\documentclass[alpha-refs, 11pt]{article}
\usepackage[margin=1in]{geometry}

\usepackage[T1]{fontenc}

\usepackage{lmodern}
\usepackage{sansmathfonts}

\usepackage[table]{xcolor}
\usepackage{ifpdf}
\usepackage{ifxetex}
\usepackage{ifluatex}
\usepackage{mathtools}
\usepackage{bm}
\usepackage{graphicx}
\usepackage{tabularx}
\usepackage{textcase}
\usepackage{dashrule}
\usepackage{ragged2e}
\usepackage{authblk}
\usepackage[hyphens]{url}
\usepackage{soul}
\usepackage{xpatch}
\usepackage{afterpage}
\usepackage{amsmath}
\usepackage{amsfonts}
\usepackage{natbib}

\usepackage{siunitx}
\usepackage{multirow}
\usepackage{hyperref}

\DeclareMathOperator*{\argmin}{arg\,min}
\usepackage[intoc]{nomencl}
\makenomenclature

\usepackage{soul}

\newcommand{\RS}[1]{\textcolor{black}{#1}}

\usepackage{lineno}

\usepackage{etoolbox}
\renewcommand\nomgroup[1]{%
  \item[\bfseries
  \ifstrequal{#1}{A}{Acronyms}{%
  \ifstrequal{#1}{F}{Functions}{%
  \ifstrequal{#1}{P}{Parameters and Variables}{%
  \ifstrequal{#1}{S}{Miscellaneous}{}}}
  }%
]}

\title{ \vspace{-1cm}Transfer Learning Bayesian Optimization to Design Competitor DNA Molecules for Use in Diagnostic Assays}

\author[1,2]{Ruby Sedgwick}
\author[1]{John P. Goertz}
\author[1,3]{Molly M. Stevens}
\author[2]{Ruth Misener}
\author[2]{Mark van der Wilk}

\affil[1]{Department of Materials, Department of Bioengineering  and Institute of Biomedical Engineering, Imperial College London, London}
\affil[2]{Department of Computing, Imperial College London, London}
\affil[3]{Department of Physiology, Anatomy and Genetics, Department of Engineering Science, and Kavli Institute for Nanoscience Discovery, University of Oxford, Oxford}

\begin{document}

\maketitle

\begin{abstract}
    \normalsize
\noindent With the rise in engineered biomolecular devices, there is an increased need for tailor-made biological sequences. Often, many similar biological sequences need to be made for a specific application meaning numerous, sometimes prohibitively expensive, lab experiments are necessary for their optimization. This paper presents a transfer learning design of experiments workflow to make this development feasible. By combining a transfer learning surrogate model with Bayesian optimization, we show how the total number of experiments can be reduced by sharing information between optimization tasks. We demonstrate the reduction in the number of experiments using data from the development of DNA competitors for use in an amplification-based diagnostic assay. We use cross-validation to compare the predictive accuracy of different transfer learning models, and then compare the performance of the models for both single objective and penalized optimization tasks. 

\end{abstract}

\section{Introduction}

Tailoring biological sequences, such as oligonucleotides or proteins, for specific applications is a common challenge in bioengineering. These engineered molecules have a variety of uses including in biosensors \citep{hua_dna-based_2022, deng_programmable_2023, goertz_competitive_2023}, medical therapeutics \citep{badeau_engineered_2018, blakney_skin_2019, ebrahimi_engineering_2023} and bio-computing \citep{siuti_synthetic_2013, qian_neural_2011, lv_biocomputing_2021}. However, development often requires expensive or time consuming experiments, meaning good experimental design is necessary to optimize the biological sequences within the experimental budget \citep{cox_theory_2000}. This also leads to better analysis, especially when there are interaction effects between input factors, which is common in biological experiments \citep{kreutz_systems_2009, politis_design_2017, papaneophytou_design_2019, fellermann_design_2019, narayanan_bioprocessing_2020, gilman_statistical_2021}. 

Iterative experimental designs have the advantage of using information from previous experiments to inform future ones. Bayesian optimization is an iterative global black box optimization strategy \citep{snoek_practical_2012, shahriari_taking_2016} which has proven effective for design of biomolecular experiments including vaccine production \citep{rosa_maximizing_2022}, antibody development \citep{khan_toward_2023}, design and manufacturing of proteins and tissues \citep{romero_navigating_2013, mehrian_maximizing_2018, narayanan_design_2021, gamble_machine_2021}, validation of molecular networks \citep{sedgwick_design_2020} and extracellular vesicle production \citep{bader_improving_2023}. In Bayesian optimization, a surrogate model, usually a Gaussian process, of the system is built using data and an acquisition function decides which data point to collect next. Gaussian processes are a powerful tool for designing biological experiments in low data regimes due to their uncertainty estimates \citep{hie_leveraging_2020}.

When many similar biological sequences need to be designed, it can be harder to optimize all the sequences within the experimental budget. Optimizing each sequence from scratch discards useful information from previous tasks, meaning more experiments are required. An alternative is to use transfer learning --- a technique that improves the learning of new sequences by using knowledge gained from other optimization tasks \citep{zhuang_comprehensive_2021}. Transfer learning is closely related to multi-task learning, where information is shared between tasks that are optimized at the same time. The approach outlined here can be used for either, and we will use transfer learning as an umbrella term for both.

As we require our surrogate model to be data efficient and have uncertainty quantification, we consider four Gaussian process models: an average Gaussian process (AvgGP), the multi-output Gaussian process (MOGP), the linear model of coregionalisation (LMC) and the latent variable multi-output Gaussian process (LVMOGP). The key difference between these Gaussian process models lies in their handling of correlations between outputs: from no correlation in the MOGP to non-linear correlation in the LVMOGP. 

\begin{figure*}[t!]
\begin{center}
    \includegraphics[width=0.9\textwidth]{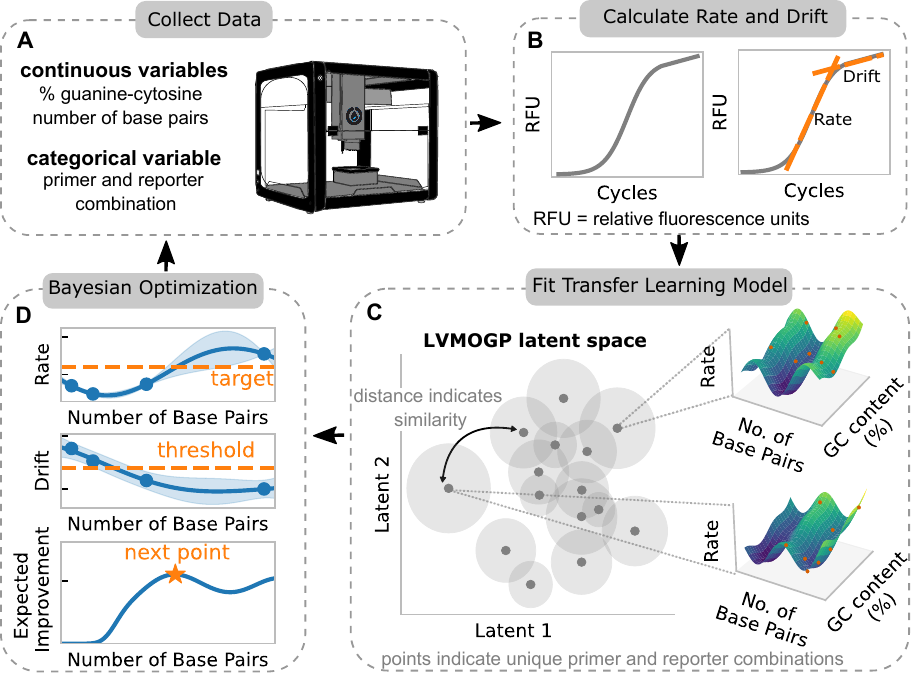}
\end{center}
\caption{Design of experiments workflow for optimizing the competitor DNA molecules. \textbf{(A)} Data is collected in the lab using a DNA amplification reaction assay. \textbf{(B)} The rate and drift are then calculated by fitting amplification curves. \textbf{(C)} A transfer learning surrogate model uses the data to predict the rate and drift for each of the given competitors. The LVMOGP is introduced in Section~\ref{sec:lvmogp}. Information is shared through the latent space, with one point on the latent space for each competitor. The shaded regions indicate the uncertainty. The 3D plots are predictions of the model for given competitors. \textbf{(D)} The Bayesian optimization algorithm, introduced in Section~\ref{sec:bayes_opt}, combines information about the rate and drift surfaces in an acquisition function to select the experiment to run for each competitor. The solid lines in the rate and drift plots represent the mean of the Gaussian process models, while the shaded regions are \(2\times\text{standard deviation}\). This process is repeated until all optimal competitor sequences are found or the experimental budget is exhausted.}
\label{fig:graphical_abstract}
\end{figure*}

We apply these surrogate models in conjunction with Bayesian optimization for efficient optimization of bio-molecules, as shown in Figure~\ref{fig:graphical_abstract}. We focus specifically on the development of a new modular diagnostic assay, based on competitive polymerase chain reaction (PCR), for measuring expression of multiple genes simultaneously, giving a single end point readout \citep{goertz_competitive_2023}. This diagnostic requires many competitor DNA sequences to be optimized to have the correct amplification properties in PCR reactions, and we believe the relationship between the responses of the competitors may be non-linear. For optimal results, these competitors should have a predefined amplification curve rate; and a nuisance drift factor should ideally be below a certain threshold to allow for a more stable readout. 

We use synthetic data experiments to compare the Gaussian process models in different settings. We then use cross-validation to verify the benefit of the LVMOGP for modeling the response of the competitors, using data from DNA amplification experiments. We confirm that a LVMOGP surrogate model in conjunction with the design of experiments workflow speeds up optimization of the competitors both when only the single objective of rate is optimized and when rate is optimized with a penalty on drift over a given threshold.

\section{Materials and Methods}

\subsection{Gaussian Process Regression}
A Gaussian process is a stochastic process representing an infinite collection of random variables, the joint distribution of any subset of which is a multi-dimensional Gaussian distribution \citep{rasmussen_gaussian_2006}.  A Gaussian process is fully defined by its mean \(m: \mathbb{R}^D \mapsto \mathbb{R}\) and covariance  \(k: \mathbb{R}^D \times \mathbb{R}^D \mapsto \mathbb{R}\) functions:
\nomenclature[F]{\(m(\cdot)\)}{Gaussian Process mean function}   
\nomenclature[F]{\(k(\cdot, \cdot)\)}{Gaussian Process covariance function} 
\nomenclature[F]{\(f(\cdot)\)}{Function of \(\mathbf{x}\)}   
\nomenclature[P]{\(\mathbf{x}\)}{Input location such that \(\mathbf{x} \in \mathbb{R}^{D}\)}  
\nomenclature[P]{\(D\)}{Dimensions of \(\mathbf{x}\)} 
\nomenclature[F]{\(\mathcal{GP}\)}{Gaussian Process} 
\begin{equation}
\label{basic_Gaussian_process}
f(\mathbf{x}) \sim \mathcal{GP}\left(m(\mathbf{x}), k(\mathbf{x}, \mathbf{x}')\right),
\end{equation}
where \(\mathbf{x}\in \mathbb{R}^D\) is our input.  For a full nomenclature see Appendix~\ref{supp:nom}. We assume our output data \(y(\mathbf{x}) \in \mathbb{R}\) to be noisy evaluations of \(f(\mathbf{x}) \in \mathbb{R}\):
\begin{equation}
\label{basic_Gaussian_process_y}
y(\mathbf{x}) = f(\mathbf{x}) + \epsilon,
\end{equation}
\nomenclature[P]{\(y\)}{Noisy evaluations of  \(\mathbf{x}\)} 
\nomenclature[P]{\(\epsilon\)}{Noise added to \(y\) where  \( \epsilon \sim \mathcal{N}(0, \sigma_n^2\mathbf{I)})\)} 
\nomenclature[P]{\(\sigma_n^2\)}{Noise variance of Gaussian process} 
where \( \epsilon \sim \mathcal{N} (0, \sigma_n^2)\) and \(\sigma_n^2\) is the noise variance.

The choice of kernel and hyperparameter initializations for a given application depends on prior information about the system, see Appendix~\ref{app:GPR} for more details. Often this implies setting the mean function to zero, which is what we do here. 
A common kernel function is the squared exponential, which is a stationary kernel that assumes the data-generating function is smooth:
\begin{equation}
\label{squared_exp}
k(\mathbf{x}, \mathbf{x'}) = \sigma_k^2 \text{exp}\left(-\sum_{d=1}^D\frac{(x_d-x_d')^2}{2\ell_d^2}\right),
\end{equation} \noindent
where \(\sigma_k^2\) is the kernel variance and \(\ell_d\) is the lengthscale of dimension \(d\) \citep[Chapter 4]{rasmussen_gaussian_2006}. Given a set of \(N\) training data \(\mathcal{D} = \{(\mathbf{x}_i, y_i)| i= 1, ..., N\}\), the training inputs \(\{x_n\}_{i=1}^N\) can be aggregated into the matrix \(X \in \mathbb{R}^{N \times D}\) and the training observations \(\{y_n\}_{i=1}^N\) aggregated into the vector \(\mathbf{y} \in \mathbb{R}^{N}\). It is then possible to write a joint distribution of the training observations \(\mathbf{y}\) and predicted function value \(\mathbf{f_*}\) at prediction locations \(X_*\). Thus, the mean and covariance of the Gaussian process at the prediction points can be calculated respectively:

\nomenclature[P]{\(\sigma_k^2\) }{Kernel variance} 
\nomenclature[P]{\(\ell_d\)}{Lengthscale of dimension \(d\)} 
\nomenclature[P]{\(X\)}{Training inputs of Gaussian Process \(X = \{\mathbf{x_1}, ..., \mathbf{x_N}\} \in \mathbb{R}^{N \times D} \)} 
\nomenclature[P]{\(X_*\)}{Locations to be evaluated} 
\nomenclature[P]{\(\mathbf{f_*}\) }{Predictions at locations \(X_*\)} 
\nomenclature[P]{\(\mu(X_*)\) }{Predicted mean at locations \(X_*\)} \nomenclature[P]{\(\sigma(X_*)\)}{Predicted covariance at locations \(X_*\)} 
\nomenclature[P]{\(\theta\)}{Gaussian Process hyperparameters} 

\begin{align}
    \label{GP_mean_prediction}
    \mu(X_*) &= \mathbb{E}[\bar{\mathbf{f}}_*|X, \mathbf{y}, X_*] \\ &= K(X_*, X)[K(X, X) + \sigma_n^2\mathbf{I}]^{-1} \mathbf{y}
\end{align}

\begin{align}
    \label{GP_var_prediction}
    \sigma(X_*)= K(X_*, X_*) - K(X_*, X)[K(X, X) + \sigma_n^2\mathbf{I}]^{-1}K(X, X_*).
\end{align}

\noindent The hyperparameters \( \mathbf{\theta} = \{\sigma_n^2, \sigma_k^2, \ell_d\}\) are optimized by maximizing the marginal likelihood \(p(\mathbf{y}|X, \mathbf{\theta})\), which is calculated in closed form \citep[Chapter 2]{rasmussen_gaussian_2006}. 

\subsection{Gaussian Processes with Multiple Outputs}

\subsubsection{Independent Gaussian Processes with Shared Kernel}

The multi-output Gaussian process (MOGP) allows for multiple outputs such that \(\mathbf{y} \in \mathbb{R}^{N \times P}\) \citep{alvarez_kernels_2012}. All outputs have the same kernel function and hyperparameters but function values on different outputs are uncorrelated. This means the kernel of the MOGP is a block diagonal with \(k(X_p, X'_p) = k(X_p, X'_p)\) if \(p=p'\) and \(k(X_p, X'_p) = 0\) if \(p\neq p'\) where \(p\) is the output index. The joint distribution for two outputs \(\mathbf{f_1}\) and \(\mathbf{f_2}\) evaluated at points \(X_1\) and \(X_2\) is given by:
\begin{equation}
    \label{eq:MOGP_kernel}
    \begin{bmatrix}
        \mathbf{f_1} \\
        \mathbf{f_2}
    \end{bmatrix} \sim \mathcal{N} \left( \mathbf{0}, \begin{bmatrix}
        K(X_1, X_1) & \mathbf{0} \\
        \mathbf{0} & K(X_2, X_2) \\
    \end{bmatrix}  \right).
\end{equation}
\nomenclature[P]{\(P\)}{Number of output functions in multi-output Gaussian Process} 
We use the MOGP to demonstrate the setting of no transfer of information about function values.

\subsubsection{Linear Model of Coregionalization}

The linear model of coregionalization (LMC) extends the MOGP to model linear correlations between output surfaces by assuming they are linear combinations of Gaussian process latent functions:
\begin{equation}
    f_p(\mathbf{x}) = \mathbf{W}_p\mathbf{g}(\mathbf{x}) + \kappa_p \upsilon_p(\mathbf{x}).
\end{equation}
\nomenclature[P]{\(\mathbf{W}\)}{Vector of weights of the latent functions in the linear model of coregionalisation}
\nomenclature[F]{\(\mathbf{g}(\cdot)\)}{Latent Gaussian processes in the linear model of coregionalisation}
\noindent where \(\mathbf{W} \in \mathbb{R}^{P\times Q}\) is a vector of weights \(\mathbf{g}(\mathbf{x}) = \{g_q(\mathbf{x})\}_{q=1}^Q\) are shared latent functions, \(\upsilon_p(\mathbf{x})\) is a latent function that allows for some independent behavior and \(\kappa_p\) is a learned constant \citep{alvarez_kernels_2012, bonilla_multi-task_2007}.

This leads to a Kronecker structured kernel such that the joint distribution between two functions \(\mathbf{f}_1\) and \(\mathbf{f}_2\) is given by:
\begin{equation}
    \label{eq:lmc_joint}
   \begin{bmatrix}
        \mathbf{f_1} \\
        \mathbf{f_2}
    \end{bmatrix} \sim \mathcal{N} \left( \mathbf{0}, \begin{bmatrix}
        \sum_{q=1}^Q b_{11}  k_q(X_1, X_1) & \sum_{q=1}^Q b_{12} k_q(X_2, X_2) \\
        \sum_{q=1}^Q b_{21} k_q(X_1, X_1) &  \sum_{q=1}^Q b_{22} k_q(X_2, X_2) \\
    \end{bmatrix}  \right),
\end{equation}

\nomenclature[P]{\(B\)}{Coregionalization matrix in the LMC} 
\nomenclature[P]{\(Q\)}{Number of covariance matrices in the LMC} 

\noindent where \(b_{pp'}\) is an element of \(B=WW^T + \text{diag}(\mathbf{\kappa})\), a \(P\times P\) matrix determining the similarity between functions and there are Q different covariance functions \(k_q(\mathbf{x}, \mathbf{x'})\).  If \(Q=1\), this is known as the intrinsic coregionalization model \citep{alvarez_kernels_2012}.

Coregionalization methods have successfully been used for Bayesian optimization \citep{cao_adaptive_2010, swersky_multi-task_2013, tighineanu_transfer_2022} and applied to the optimization of synthetic genes \citep{gonzalez_bayesian_2015} and chemical reactions \citep{taylor_accelerated_2023}. However, coregionalization methods assume the response surfaces are linear combinations of a small number of latent functions, so they can fail to fit and predict well on data with non-linear similarity between surfaces. 

\subsubsection{Latent Variable Multi-output Gaussian Process} \label{sec:lvmogp}

 The latent variable multi-output Gaussian process (LVMOGP) introduced by \cite{dai_efficient_2017} can model non-linear similarities. It does so by augmenting the input domain of a Gaussian process with a \(Q_H\) dimensional latent space \(\mathcal{H}\). Each output function has a latent variable, such that the latent variables are denoted by \(H = [\mathbf{h}_1, ..., \mathbf{h}_P]^T \in \mathbb{R}^{P \times Q_H}\). The LVMOGP assumes output \(\mathbf{y}_p\) is generated by:
 \nomenclature[S]{\(\mathcal{H}\)}{The latent space in the LVMOGP} 
 \nomenclature[P]{\(\mathbf{h_p}\)}{Latent variable of the \(p^{th}\) output function} 
 \nomenclature[P]{\(H\)}{Latent variables such that  \(H = [\mathbf{h_1}, ..., \mathbf{h_P}]^T \in \mathbb{R}^{Q_H \times P}\)} 
\nomenclature[P]{\(\mathbf{I}\)}{Identity matrix} 
\begin{equation}
    \label{eq:LVMOGP_y}
    y_p(\mathbf{x}) = f(\mathbf{x}, \mathbf{h_p}) + \epsilon,
\end{equation}
where  \(\epsilon \sim \mathcal{N}(0, \sigma_n^2\mathbf{I})\). The latent space allows the LVMOGP to automatically transfer learn between output functions as it will cluster similar output functions together and place wildly different ones far apart on the latent space. The distance in the latent space and the latent space lengthscale determines the amount of correlation between different output functions. To account for uncertainty in the placement of the latent variables, they are treated as distributions rather than point estimates, such that \(\mathbf{h_p} \sim \mathcal{N}(\mathbf{\mu}_{h_p}, \Sigma_{h_p})\). For more details on the implementation of the LVMOGP see Appendix~\ref{supp:LVMOGP_VI}. 
 \nomenclature[P]{\(\mathbf{\mu}_{h_p}\)}{Mean of the \(p^{th}\) latent variable}
 \nomenclature[P]{\(\Sigma_{h_p}\)}{Variance of the \(p^{th}\) latent variable}

Similar latent variable models have been used for Bayesian optimization of material development \citep{zhang_latent_2020} and for transfer learning across cell lines \citep{hutter_knowledge_2021}. However, these methods treat the latent variables as point estimates rather than distributions as in the LVMOGP, which can cause poor uncertainty estimates, especially at low data regimes.

\subsubsection{Comparison of Gaussian Process Models}

\begin{figure*}[t]
\centering
\includegraphics[width=\textwidth]{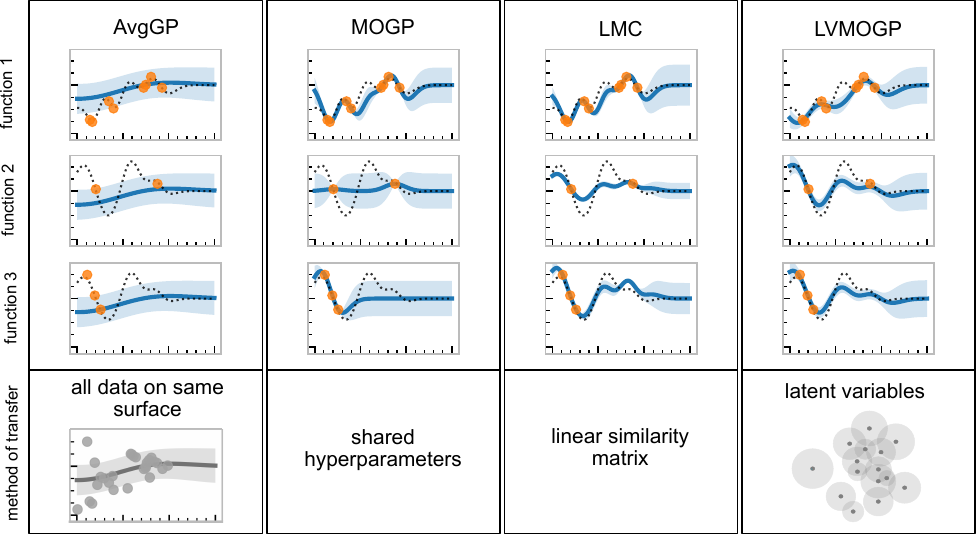}

\caption{Predictions of the four Gaussian process models fitted to a toy dataset with linear correlation between output surfaces. The dots are the data, the dashed line is the true function, the solid line is the Gaussian process mean prediction and the shaded region is two times the predicted standard deviation, meaning around \(95\%\) of the data points should lie within the shaded region. The bottom row explains how data is transferred between the surfaces by each model. For the average Gaussian process (AvgGP), all data is assumed to be from the same surface, for the multioutput Gaussian process (MOGP) information is only transferred about the hyperparameter values but not the function values. In the linear model of coregionalisation (LMC) information is transferred via the similarity matrix \(\mathbf{B}\) and in the latent variable multiouput Gaussian process (LVMOGP) it is transferred through the latent space. Theoretically, LMC and LVMOGP can learn if information can be transferred and (if so), how much.}
\label{fig:model_comparison}
\end{figure*}

In our comparisons, we include a fourth model called the average Gaussian process (AvgGP), which treats all the data as if it has come from the same response surface. Figure~\ref{fig:model_comparison} shows predictions of the four Gaussian process models on a toy data set with linear correlation between output functions. See Appendix ~\ref{app:toy_dataset} for details of the data generation. As the AvgGP doesn't differentiate between surfaces, it doesn't fit any response surface well. The MOGP only shares hyperparameters but no information about function values between response surfaces, meaning it makes worse predictions and has more uncertainty on new response surfaces. The LMC has a better mean prediction than the MOGP as it shares information between response surfaces. The LVMOGP similarly has better mean prediction than the MOGP as it shares information across response surfaces through the latent space. If \(Q=1\) and \(B\) is the identity matrix, then the LMC recovers the MOGP. If a linear kernel is applied to the latent dimensions of the LVMOGP, the LMC is recovered, and by making the distance between latent variables large relative to the lengthscale, the MOGP can be recovered too. The fact there are hyperparameter settings for the LMC and LVMOGP that recover the MOGP is promising for preventing negative transfer, as in the case where there is no correlation between response surfaces they can just revert to the MOGP. However, this is only true for large data sets --- in low data regimes, we may expect some negative transfer in the no correlation case, due to uncertainty in the hyperparameter values and, in the case of the LVMOGP, a prior on the existence of correlations. 

\subsubsection{Gaussian Process Implementation}

Details of the data processing and hyperparameter initialisations can be found in Appendix~\ref{app:GPR}. All coding was done in Python using version 3.9.  The Gaussian process models were implemented using GPFlow 2.3.0 \citep{matthews_gpflow_2017}. GPFlow has implementations of the standard Gaussian process, MOGP and the LMC. Our LVMOGP was implemented as a new GPflow model class, which can be accessed via the Github links in Appendix~\ref{supp:github}. Other packages used include PyMC3 3.11.4 \citep{salvatier_probabilistic_2016} for Bayesian parameter estimation, Numpy 1.21.4 \citep{harris_array_2020}, Scipy 1.7.1 \citep{virtanen_scipy_2020} and Pandas 1.3.4 \citep{the_pandas_development_team_pandas-devpandas_2023} for data processing and Matplotlib 3.4.3 \citep{droettboom_matplotlib_2015} for visualization. 

\subsection{Bayesian Optimization} \label{sec:bayes_opt}

Bayesian optimization is a sequential experimental design strategy for finding the global minimum (or maximum) of an objective function \citep{shahriari_taking_2016, snoek_practical_2012}. As the objective function is unknown, a surrogate model is used to represent the posterior belief of the objective function and updated every time a new data point is observed. An acquisition function is then used to select the next data point to collect. A common acquisition function is the expected improvement which trades off exploration of regions with little data and exploitation of regions which are expected to be optimal \citep{jones_efficient_1998, garnett_bayesian_2023}. This process is repeated until the optimum has been found or the experimental budget is exhausted.

\subsubsection{Acquisition Function}

Rather than maximizing or minimizing the rate, as is usual in Bayesian optimization, we wish to minimize the difference between the rate, \(f_{\text{rate}}\), and the target rate, \(T_{\text{rate}}\): 
\begin{equation}
    \argmin_{BP, GC} ~~ \sqrt{(f_{\text{rate}} - T_{\text{rate}})^2} 
\end{equation}
\nomenclature[F]{\(f_{\text{rate}}(\cdot)\)}{Rate function in competitor amplification}
\nomenclature[P]{\(T_{\text{rate}}\)}{Target rate}
\noindent Therefore, we use the target vector optimization acquisition function, that extends the expected improvement acquisition function to minimize the Euclidean distance between a target vector and a vector of the current predicted values \citep{uhrenholt_efficient_2019}. As we are only optimizing the rate, we use their formulation with scalars instead of vectors. In this formulation, a stochastic variable is defined as \(\mathbf{\delta}|\mathbf{x}=\left\lVert y_{\text{rate}}(\mathbf{x}) - T_{\text{rate}} \right\rVert_2^2\) where \(y_{\text{rate}}(\mathbf{x})\) is the output value at input \(\mathbf{x}\) and \(T_{\text{rate}}\) is our target value. The distribution of \(p(\mathbf{\delta}|\mathbf{x})\) is modeled with the aim of minimizing \(\delta\). If the response surfaces are Gaussian processes, then \(p(\mathbf{\delta}|\mathbf{x})\) can be approximated using a non-central \(\chi^2\) distribution \citep{uhrenholt_efficient_2019}. The expected improvement for this non-central \(\chi^2\) distribution is expressed as:
\nomenclature[F]{\(\alpha_{EI}(\cdot)\)}{Expected improvement acquisition function} 
\nomenclature[P]{\(\lambda\)}{Non-centrality parameter of target vector optimization expected improvement} 
\nomenclature[P]{\(t\)}{\(=\delta \gamma^{-2}\)} 
\nomenclature[S]{\(G\)}{An approximation to the cumulative non-central \(\chi^2\) distribution function} 
\nomenclature[P]{\(\mathbf{\delta}\)}{stochastic variable defined as the squared difference between observed outputs and the target value}
\begin{align}
\label{eq:TVBO_EI}
    \alpha_{EI} = \mathbf{\delta}_{min} G_{ \lambda}(\mathbf{\delta}_{min}/\gamma^2)  - \gamma^2 \mathbb{E}[t|t < \mathbf{\delta}_{min}/\gamma^2]G_{\lambda}(\mathbf{\delta}_{min}/\gamma^2),
\end{align}
where \(\mathbf{\delta}_{min}\) is the minimum \(\mathbf{\delta}\) observed so far, \(\gamma\) is root mean of the variances of each output evaluated at the training points, \(t=\mathbf{\delta} \gamma^{-2}\), and \(G_{\lambda}\) is an approximate cumulative \(\chi^2\) distribution with non-centrality parameter \(\lambda\) defined in the paper \citep{uhrenholt_efficient_2019}.

\subsubsection{Bayesian Optimization with Drift Penalty}

To ensure the drift value remains below, or close to the threshold, we use the \emph{probability of feasibility} to encourage the algorithm to select points that have a high chance of being below the threshold \citep{schonlau_global_1998}:
\begin{equation}
    PF(\mathbf{x}) = p(f_{\text{drift}}(\mathbf{x}) \leq  T_{\text{drift}}),
    \label{eq:PF}
\end{equation}
where \(f_{\text{drift}}(\mathbf{x})\) is the value of drift function at \(\mathbf{x}\), and \(T_{\text{drift}}\) is the drift threshold. 
\nomenclature[F]{\(PF(\cdot)\)}{Probability of feasibility}
\nomenclature[F]{\(\alpha_{c}(\cdot)\)}{Acquisition function including probability of feasibility}
We then multiply the expected improvement by the probability of feasibility to get our final acquisition function:
\begin{equation}
    \alpha_c = PF(\mathbf{x})\alpha_{EI}(\mathbf{x}).
    \label{eq:constrained_acq}
\end{equation}
The probability of feasibility has been used for optimization applications including analog circuits \citep{lyu_efficient_2018} and materials design \citep{sharpe_design_2018}.

\subsubsection{Performance Metrics}

For both the synthetic experiments and the cross-validation experiments we assessed the fit of Gaussian process models with two performance metrics: root mean squared error (RMSE):
\begin{equation}
    \text{RMSE} =\sqrt{\frac{\sum^{N^*}_{i=1}(\mu(\mathbf{x}^*_i) - y^*_i)^2}{N^*}},
    \label{eq:rmse}
\end{equation}
 and negative log predictive density (NLPD):
 \begin{align}
    \text{NLPD} &= \frac{1}{N^*} \sum^{N^*}_{i=1} \log p(y_i^* |\mathbf{x}^*_i, \mathbf{X}, \mathbf{y}, \mathbf{\theta}) \\ &=- \frac{1}{2N^*} \sum^{N^*}_{i=1} \left( -\log(2\pi\sigma(\mathbf{x}^*_i)^2) -\frac{\left(y^*_i - \mu(\mathbf{x}^*_i)\right)^2}{\sigma(\mathbf{x}^*_i)^2}\right).
    \label{eq:nlpd}
 \end{align}

\noindent These are both calculated on a test set of input locations \(X^*\) of length \(N^*\). The RMSE is useful for comparing the mean predictions of the Gaussian processes, while the NLPD also indicates how good the uncertainty estimate is, both of which are important for effective exploration and exploitation.
For assessing the Bayesian optimization algorithm, we use cumulative regret:

\begin{align}
    \text{regret} = \min_{i\in[1..N]} \left( \sqrt{y_{\text{rate},i} - y_{\text{best}})^2} \right. \left. + \max(0, y_{\text{drift},i} - T_{\text{drift}})\vphantom{\sqrt{\mu(\mathbf{x}_{*i}) - y_{\text{best}})^2}}\right),
    \label{eq:regret}
\end{align}
where \(y_{\text{rate},i}\) and \(y_{\text{drift},i}\) are rate and drift training data, \(y_{best}\) is the data point closest to the target out of both training and candidate sets for that surface and \(\max(0, (y_{\text{drift},i} - T_{\text{drift}})\) is a penalty for exceeding the drift threshold.

\nomenclature[P]{\(y_{best}\)}{Data point which is closest to the target out of the train and test datasets for a given surface} 
\nomenclature[F]{\(f_{\text{drift}}(\cdot)\)}{Drift function} 
\nomenclature[P]{\(\mathbf{y}_{\text{drift}}\)}{Drift output data} 
\nomenclature[P]{\(\mathbf{y}_{\text{rate}}\)}{Rate output data} 
\nomenclature[P]{\(T_{drift}\)}{Drift threshold} 

\subsection{Data Collection}

\begin{figure}[bt]
\centering
\includegraphics[width=8cm]{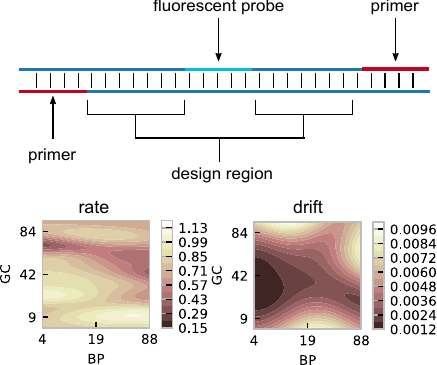}
\caption{Schematic of the competitor design space. For a given competitor DNA molecule, the primers and fluorescent probe regions are fixed. We can edit the design region to ensure the sequence has a given number of base pairs and guanine-cytosine content. Changing the number of base pairs and guanine-cytosine-content affects the rate and drift of the competitor, allowing us to fine-tune to the rate and drift required for the diagnostic assay.}
\label{fig:cPCR}
\end{figure}

Each competitor has predefined primers and fluorescent probes and a design region where the sequence can be altered. Rather than tackling the difficult combinatorial problem of optimizing the sequence directly, we reduce the problem to two key input variables: the number of base pairs (BP) and guanine-cytosine content (GC) as in Figure~\ref{fig:cPCR}. This converts the design space into a more manageable continuous form and reduces the input dimensions, which is beneficial when data is limited. For each BP-GC combination, chosen by an expert researcher, a polymerase chain reaction (PCR) assay generates an amplification curve, from which rate and drift are calculated. In total, we have data on 34 different competitors and wish to optimize 16 of these. Across the 34 competitors, we have 592 data points at 327 unique input locations, with 1 to 6 repeats at each location. 
See Appendix~\ref{data_summary} for a summary of the data.

The rate and drift for each amplification curve were calculated using the following equations: 
\begin{equation}
    F_T = \frac{\nu}{1 + \frac{(\nu-F_0)}{F_0} \cdot e^{-r\cdot \tau}},
     \label{eq:F}
\end{equation}

\begin{equation}
\text{signal} = F_T \cdot \left(1 + \frac{F_T}{\nu} \cdot m \cdot \left(\ln(F_0)/r\right)\right), 
     \label{eq:rate_drift_fitting}
\end{equation}
where \(F_T\) and \(F_0\) are the end point and starting fluorescence, \(\nu\) is carrying capacity, \(r\) is the rate, \(m\) is the drift and \(\tau\) is cycle number. 
\nomenclature[P]{\(F_0\)}{Fluorescence at the beginning of the DNA amplification reaction}
\nomenclature[P]{\(F\)}{Fluorescence in DNA amplification reaction}
\nomenclature[P]{\(F_T\)}{Fluorescence at the end of the DNA amplification reaction}
\nomenclature[P]{\(\nu\)}{Carrying capacity}
\nomenclature[P]{\(\tau\)}{Cycle number}

\subsubsection{Polymerase Chain Reactions}

To perform the PCR reactions, we used an Applied Biosystems QuantStudio 6 Flex using Applied Biosystems MicroAmp EnduraPlate Optical 384-well plates (Thermo Fisher Scientific, Waltham, MA, USA). The theromcycling stages consisted of a melt step at 95\textdegree C for 3 seconds and an annealing step at 60\textdegree C. All reactions were performed at 10 \textmu L and used Applied Biosystems TaqMan Fast Advanced Master Mix. Either fluorescent probes or EvaGreen dye (Biotium, Fremont, CA, USA) were used as reporters. 

\subsubsection{DNA Sequences}

For each BP-GC combination for a given competitor, NUPACK \citep{zadeh_nupack_2011} was used to create a DNA sequence with the correct number of base pairs and guanine-cytosine content, as well as the correct sequences for the primer and probes. These sequences, alongside synthetic natural target analogs, were purchased from Twist Biosciences (San Francisco, CA) or as eBlock Gene Fragments from Integrated DNA Technologies (“IDT”, Coralville, IA, USA). Primers and probes were also purchased from IDT.

\section{Results}

\subsection{Synthetic Data Experiments} \label{section:synth_res}

\begin{figure*}[hbt]
\centering
\includegraphics[width=\textwidth]
{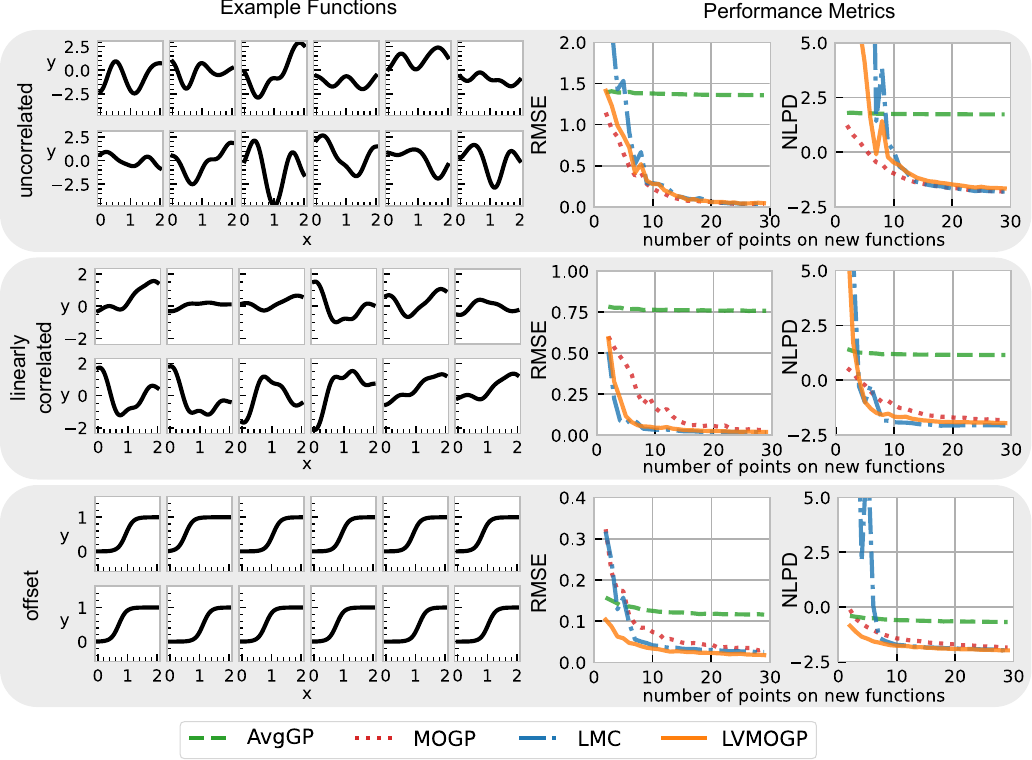}
\caption{Results of experiments with synthetically-generated data. The plots on the left show example data-generating functions used for the synthetic experiments. The plots on the right show the RMSE and NLPD for the three different test response surface types for each of the Gaussian process models. New points are added randomly, and each line is the mean of 5 different randomly generated data sets, all generated from the same test functions.}
\label{fig:synth_experiments}
\end{figure*}

To explore the performance of the MOGP, AvgGP, LMC and LVMOGP, we ran experiments on synthetic data sets representing three test cases: uncorrelated, linearly correlated and horizontally offset response surfaces. All synthetic experiments had two response surfaces each with 30 points observed and 10 new response surfaces with no points observed initially. We added one random point to each new response surface every iteration and recorded the RMSE and NLPD for the Gaussian process models' predictions. Figure \ref{fig:synth_experiments} shows the RMSEs and NLPDs of the Gaussian process models for these test settings. 

For the uncorrelated test case, response surfaces were generated as independent samples of a Gaussian process prior with a \(\ell=0.3\) and \(\sigma_k^2=2\). This test case was to check for negative transfer, where the sharing of information hinders rather than aids the learning process. In Figure \ref{fig:synth_experiments}, for the uncorrelated case the MOGP outperforms the other Gaussian process models for RMSE and NLPD until approximately 10 data points, although the RMSE of the MOGP at this point is still high. We expect the LMC and LVMOGP to have some negative transfer at very low data regimes as they have a prior expectation of correlations between response surfaces. However, with enough data, they should perform the same at the MOGP, which is corroborated by the results in Figure \ref{fig:synth_experiments}. Specifically, once the MOGP gets a reasonably low RMSE of \(<0.25\) the LMC and LVMOGP have achieved similar performance.

The response surfaces for the linearly-correlated test case were created as linear combinations of two latent functions, both generated as independent samples of a Gaussian process with squared exponential kernel, Equation~\ref{squared_exp}, and \(\ell =0.3\) and \(\sigma_k^2 = 2\). The LMC outperforms the other two Gaussian process models except at very low data regimes, which is likely due to overconfidence of the LMC when it has little data. The LMC and LVMOGP outperform the MOGP even at high data regimes, showing the advantage of transfer learning.

The horizontally offset test case was chosen as a simple example where the LMC struggles to fit the data. The response surfaces were generated by offsetting a sigmoid function horizontally by a random constant. In this case, the LVMOGP outperforms the other Gaussian process models for both RMSE and NLPD. This is because the LVMOGP can learn new surfaces with very few data points, as all it needs to do is to correctly predict where the sloped region is. The LMC performs worse than the LVMOGP because the offset cannot be represented by a linear combination of its latent functions, meaning it requires more data to perform as well.

Across all the test cases, the LMC has poor NLPD at low data regimes. This is likely because it cannot express uncertainty in the deterministic \(\mathbf{B}\) matrix.

\subsection{Prediction of DNA Amplification Experiments} \label{sec:cross_val}

The performance of the proposed design of experiments workflow was validated using data from competitor DNA amplification experiments. This was done in three parts: first cross-validation was performed to compare the predictive accuracy of the Gaussian process models; then a Bayesian optimization procedure was used to optimize only the rate; finally the Bayesian optimization with drift penalty procedure was applied.

\begin{figure*}[t]
\centering
\includegraphics[width=13cm]{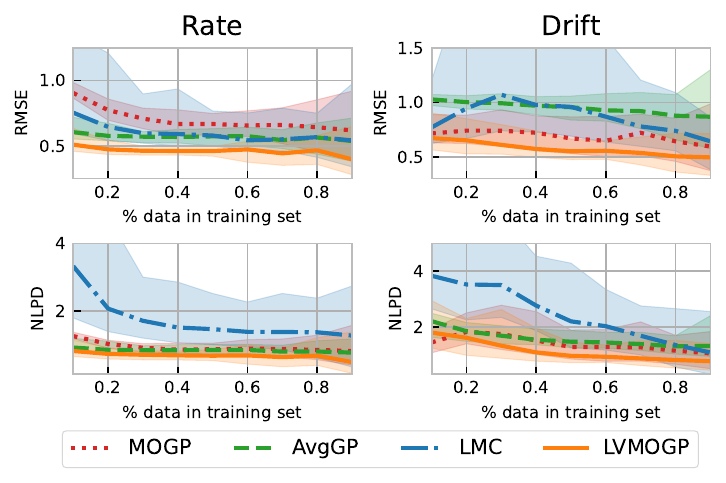}
\caption{Results of cross-validation on the DNA amplification data for both rate and drift. For each cross-validation run, the training set consisted of all the data from two competitors and a random subset of the data on the remaining competitors, ensuring all competitors had at least one data point. This is repeated for different percentages of data in the training set, and for each percentage, it is repeated 70 times.}
\label{fig:cross_validation}
\end{figure*}

In cross-validation, the training set consisted of all the data from the two competitors that had the most observations as well as a random subset of the remaining data, but ensuring all competitors had at least one data point. This was repeated 70 times for each percentage of data in the training set. We set both the rank of \(\mathbf{B}\) for the LMC and the latent dimensions of the LVMOGP to \(10\), see Appendix~\ref{app:latent_dims} for a discussion on setting these parameters. Figure \ref{fig:cross_validation} shows the RMSE and NLPD of the Gaussian process models' predictions. The LVMOGP outperforms the other Gaussian process models for both RMSE and NLPD for both rate and drift. The LMC has poor NLPD in comparison to the other Gaussian process models, suggesting it has poor uncertainty estimates.

The AvgGP model shows little improvement with increased amounts of training data. This shows the limitations of averaging the surfaces and justifies modeling each response surface separately. 

\subsection{Optimization of DNA Amplification Experiments} \label{sec:bayes_opt_res}

Ideally, for the Bayesian optimization experiments we would integrate the algorithm into the experimental loop, collecting new data with each new recommendation of each Gaussian process model. However, due to the cost of experiments, this was infeasible. Instead, we performed retrospective Bayesian optimization using the existing competitive DNA amplification dataset. The data was split into training and candidate sets, with the design of experiments algorithm only allowed to choose the next point out of the candidate set. 
Bayesian optimization was run iteratively until all points had been selected or up to a maximum number of iterations, whichever happened first.

Two learning scenarios were tested: the \textit{learning many} scenario where all data from the two competitors with the most data were fully observed to begin with and then 16 competitors optimized in parallel; and the \textit{one at a time} where each of the 16 competitors was optimized individually, with the 33 remaining competitors included in the training set. For a discussion of the effect of the choice of initial surfaces, see Appendix~\ref{app:init_surfaces}. These scenarios replicate likely lab experimentation scenarios --- the first for when many competitors need to be optimized at once, and the second for when many competitors are already optimized and an extra one is added. The maximum number of iterations was \(15\) for the rate-only optimization and \(20\) or \(10\) for the penalized optimization, depending on the learning scenario.

We also considered two methods for choosing the first experiment for a new competitor with no previously observed data. Choosing the most central data point (\textit{center} in Figure~\ref{fig:single_obj_bayes_opt}) offers both maximum reduction in variance across the response surface and ensures all competitor response surfaces have a comparable point, which may help the transfer learning methods determine their similarities. It is also a reasonable approximation of what a human experimenter without prior knowledge of the response surface might do. The second method is to let the Gaussian process model choose the first point (\textit{model's choice} in Figure~\ref{fig:single_obj_bayes_opt}) for a new competitor. For the AvgGP and the LVMOGP, this is possible as they can make posterior predictions on new response surface. For the LVMOGP, the latent variable of the new surface is determined as a weighted average of the latent variables of the response surfaces with data that have the same probe and at least one matching primer. If there are no surfaces with matching primers, we use a weighted average of the surfaces with the same probe. For the LMC and MOGP we have no posterior, so the first point is selected randomly. For these experiments, we set the values of the latent hyperparameters for the LMC and LVMOGP to \(2\), as discussed in Appendix~\ref{app:latent_dims}.

\subsubsection{Single Objective Bayesian Optimization}

\begin{figure*}[t!]
\centering
\includegraphics[width=\textwidth]{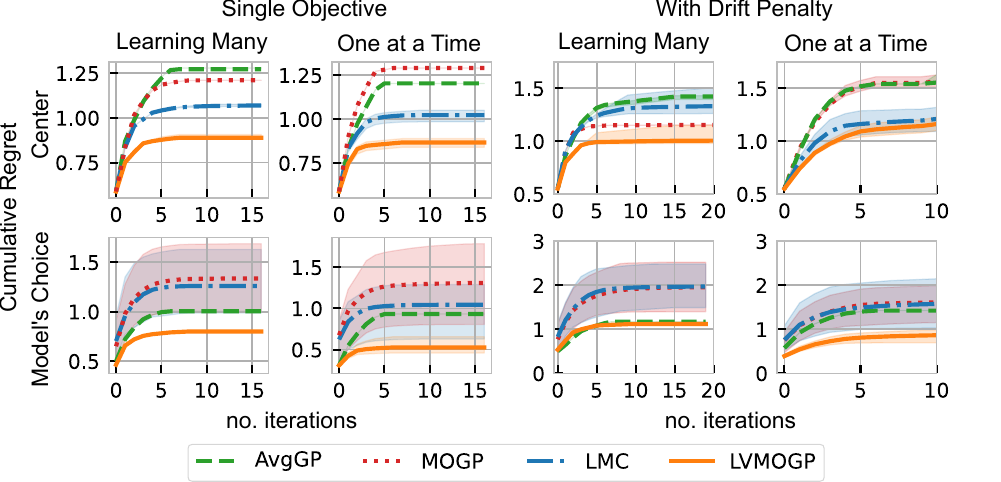}
\caption{Cumulative regret of each of the Gaussian process models for single objective (left) and penalized (right) Bayesian optimization. Each line indicates the mean across 24 random seeds and all competitors, while the shaded regions indicate the upper and lower 5\% quantiles by random seed. The top row is when the first point on each new surface is selected as being the center point, and the bottom is when the model is allowed to choose the first point. The \textit{learning many} scenario is when many competitors are being optimized at the same time, and the \textit{one at a time} scenario is when one competitor is being optimized, with all others being in the training set. }
\label{fig:single_obj_bayes_opt}
\end{figure*}

The left panel of Figure~\ref{fig:single_obj_bayes_opt} shows the results of optimizing rate without considering the drift penalty. The variance in the results comes from three sources. The first is the random selection of the next point when two points have the same expected improvement --- this causes unavoidable variation. The second is due to the Gaussian process models optimizing to different hyperparameter values due to different initializations. The different values arise because the optimization of the non-convex hyperparameter loss surfaces is difficult. The final source of variation is the random starting point for the MOGP and LMC.

In all cases, Figure~\ref{fig:single_obj_bayes_opt} shows the LVMOGP has much lower cumulative regret than the other models. The LVMOGP also reaches the best point first the most often: \(808\) times across all competitors, all learning scenarios and seeds compared to \(457\), \(498\) and \(484\) for the MOGP, AvgGP and LMC respectively. See Table~\ref{tab:best_point_unconstrained} in Appendix~\ref{app:single_obj} for a breakdown of these results. The \textit{center} start point allows us to compare the performance of the Gaussian process models without being skewed by the first point. In this case, the LMC and LVMOGP have the lowest cumulative regret\RS{, with mean values of \(1.08\) and \(0.91\) respectively at the end of optimization, compared to \(1.21\) and \(1.28\) for the MOGP and AvgGP for the \textit{learning many} case}. The ordering changes between the \textit{center} and \textit{model's choice} starting points, as in the latter the AvgGP and the LVMOGP are able to predict on new surfaces, giving them an advantage over the LMC and the MOGP when choosing the first point. For example, in \textit{learning many} \textit{model's choice} scenario, the mean regret of the first points selected by the LVMOGP and the AvgGP are \(0.464\) and \(0.499\) respectively compared to \(0.651\) and \(0.703\) for the MOGP and the LMC. Table~\ref{tab:first_points} in Appendix~\ref{supp:first_point} lists the mean regrets of the first points.

As the \textit{one at a time} scenario includes the data from all other competitors, the Gaussian process models start with far more data than the \textit{learning many} scenario. This means the AvgGP, the LMC and the LVMOGP all have less regret in the \textit{one at a time} scenario, as they are able to transfer information about the function values of competitors to improve prediction of the target competitor behavior. This is most notable for the \textit{model's choice} start point, where the AvgGP, LMC and LVMOGP have final cumulative regrets of \(0.93\), \(1.29\) and \(0.66\) respectively, compared to \(1.00\), \(1.63\) and \(0.80\) for the \textit{learning many} scenario. The MOGP does not transfer information about function values, so performs relatively worse than the other models for the \textit{one at a time} scenario, with a final cumulative regret of \(1.78\) for the \textit{one at a time} scenario as opposed to \(1.69\) for the \textit{learning many} scenario.

Across all learning scenarios and start points, the LVMOGP has the smallest mean number of iterations to get within a tolerance of \(0.05\) of the value of the best point, with the LVMOGP taking a mean of \(2.25\) iterations, while the AvgGP, MOGP and LMC take \(2.89\), \(3.02\) and \(2.93\) respectively. For \(16\) competitors, this equates to \(36\) experiments needed for the LVMOGP compared to \(49\) for the MOGP. See Appendix~\ref{app:single_obj} for a break down by learning scenario and starting point and Appendix~\ref{app:no_experiments} for box plots of the number of experiments taken by each model. The tolerance was set at \(0.05\) as this is approximately the level of experimental measurement uncertainty in the lab experiments \cite{goertz_competitive_2023}.

\subsubsection{Bayesian Optimization with Drift Penalty}

\begin{figure*}[thb]
\centering
\includegraphics[width=\textwidth]
{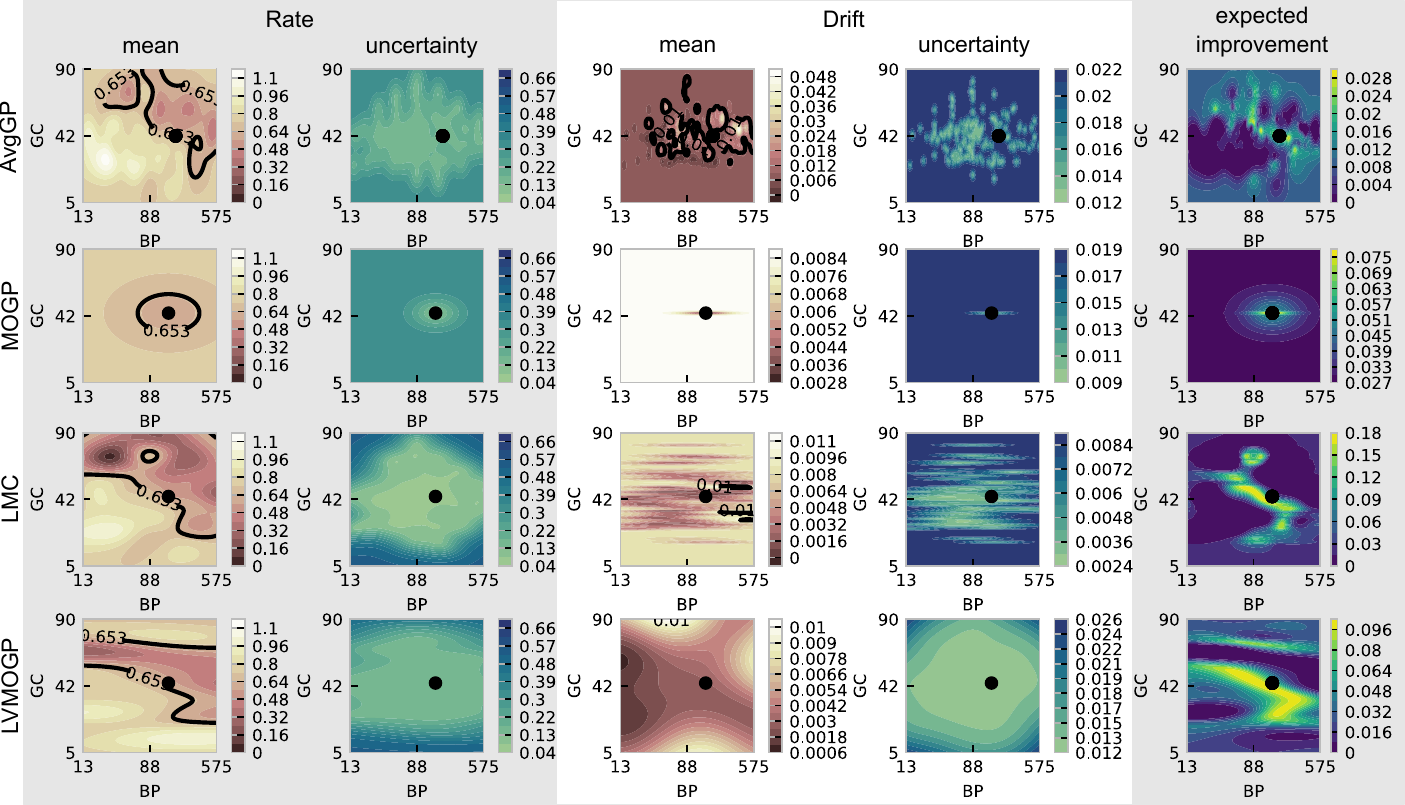}
\caption{Predictions for the rate and drift for each of the Gaussian process models. The BP and GC axes are in log and logit scales respectively. These plots show the mean of the Gaussian process model predictions and the uncertainty which here is \(2\times \text{standard deviation}\). The expected improvement with probability of feasibility is then plotted in the final column. This is for the case where we are optimizing competitor FP005-FP004-EvaGreen and have observed one data point so far, with the models able to choose the first point. The black contour lines on the mean plots indicate the target rate and threshold drift values.}
\label{fig:predictions}
\end{figure*}

\begin{figure}[htb]
\centering
\includegraphics[width=0.5\columnwidth]
{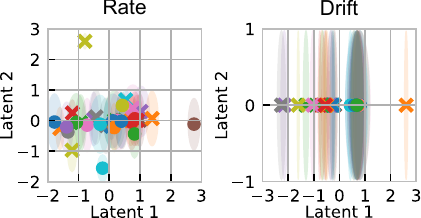}
\caption{Latent space of the LVMOGP for the rate and drift. The crosses indicate competitors with probe primers and the dots indicate those with EvaGreen primers. The shaded circles indicate the uncertainty in the latent positions.}
\label{fig:LVMOGP_latent_space}
\end{figure}

The right-hand panel of Figure~\ref{fig:single_obj_bayes_opt} shows the cumulative regret for optimization of the rate with a penalty on the drift. The LVMOGP has the lowest cumulative regret at the end for all scenarios, but doesn't outperform the other models as much as in the single objective case. In all scenarios, the MOGP, LMC and LVMOGP fail to reach the best point for the same competitor, meaning the cumulative regret curves for these models don't completely plateau. This is because they overestimate the value of the drift at the best point, so avoid selecting it. The AvgGP does find the best point for all competitors.

The LVMOGP barely outperforms the AvgGP for the \textit{learning many} scenario with \textit{model's choice} starting point.
This may be due to negative transfer in the drift predictions at very low data regimes making the selection of the first point sub-optimal. 

Similar to the single objective case, the LVMOGP has the smallest mean number of iterations to get within a tolerance of \(0.05\) of the value of the best point, with a mean of \(2.26\) iterations compared to \(3.38\), \(3.54\) and \(3.39\) for the AvgGP, the MOGP and the LMC. For \(16\) competitors, this equates to \(37\) experiments needed for the LVMOGP compared to \(57\) for the MOGP, see Appendix~\ref{app:no_experiments} for more details. Appendix~\ref{supp:bayes_opt} contains further Bayesian optimization results for both the single objective and penalized optimizations.

Figure~\ref{fig:predictions} shows the rate and drift predictions and expected improvement for one iteration. Most notably, the MOGP has no transfer of information, so has almost equal expected improvement for most of the candidate points. The other three models transfer information across the competitors, meaning even with one data point, they have much more complex predictions than the MOGP. We can also see how the AvgGP, MOGP and LMC fit the drift poorly. This is because the drift is of a different order of magnitude depending on the fluorescent probe used. Most of the Gaussian process models are unable to detect this, meaning they end up with a poor fit to the data. 

\section{Discussion}

Expensive and time consuming experiments require an intelligent design of experiments strategy. This study demonstrates how a transfer learning surrogate model can be used in conjunction with Bayesian optimization to optimize biological sequences. For the specific case of designing competitor DNA molecules for a new diagnostic, reducing the number and therefore cost of experiments can help it reach the affordability criteria for point of care settings \citep{land_reassured_2019}. 

In Bayesian optimization, we need a surrogate function with reliable mean and uncertainty estimates to ensure a balance between exploration and exploitation when selecting new points. Our cross-validation results in Section.~\ref{sec:cross_val} show the LVMOGP has better predictive accuracy than the other Gaussian process models for both rate and drift. These results also demonstrate one of the limitations of the LMC: the LMC has very high NLPD at low data regimes. This implies the LMC has poor uncertainty estimates and is overfitting, a result which has been previously observed \citep{dai_efficient_2017}.

To replicate a real-life iterative design of experiments regime, we performed Bayesian optimization on DNA amplification experimental data, but only allowing the models to select new points from existing data. For the single objective optimization case, the LVMOGP has lower cumulative regret than the other Gaussian process models for all test cases and starting points and requires fewer experiments on average to get within \(0.05\) tolerance of the best point. Specifically, the LVMOGP requires \(13\) and \(20\) fewer experiments than the no transfer MOGP model for the single objective and penalized cases respectively. This shows the LVMOGP transfer learning approach is useful both when optimizing multiple competitors at a time, and when using the data from all previous competitors to optimize a new one.

These results also demonstrate the advantage of a surrogate model that can predict unseen surfaces --- both the LVMOGP and the AvgGP see a large improvement in regret when they are allowed to select the first point, both outperforming the MOGP and LMC where the first point is chosen at random. 

When optimizing new biological sequences, there are often factors we wish to keep within a certain range such as purity \citep{degerman_constrained_2006} or biophysical properties \citep{khan_toward_2023}. While these can be treated as constraints, sometimes we may be willing to violate them slightly if it leads to a large improvement in the objective function. In these scenarios, we can add a penalty. To apply a penalty on the nuisance drift factor, we used the probability of feasibility to penalize any point predicted to be above the threshold drift value. In the penalized optimization, the LVMOGP had less cumulative regret than the other models but the difference in performance was smaller than that of the single objective optimization. This could be due to the added challenge of dealing with the penalty on drift. 

 There is variation in the performance of the Gaussian process models across random seeds due to the hyperparameter initialization. The LVMOGP has more variation due to its training being a harder optimization problem. While smart initialization and random restarts helped with this issue, future work could simplify the optimization procedures. The optimization of the Gaussian process models is discussed in Appendix~\ref{supp:initalizations}. 

While the workflow outlined here will be useful for the optimization of new competitor DNA molecules, it is not specific to this application and could be used for other applications where it is necessary to optimize many similar tasks, such as engineering DNA probes \citep{lopez_molecular_2018, wadle_real-time_2016}, optimizing conditions for different cell lines \citep{hutter_knowledge_2021}, inferring psuedotime for cellular processes \citep{campbell_bayesian_2015} or exploring protein fitness landscapes \citep{hu_protein_2023}. We expect this method will scale well to settings with more output surfaces, and predictions will improve with more data. However, as with most Bayesian optimization approaches, it will not scale as well to high dimensional input settings \citep{wang_recent_2023}.

We opted to use the LVMOGP to demonstrate how we can transfer information between tasks using proximity in latent space as Gaussian processes are data efficient and give good uncertainty predictions. However, we could replace Gaussian processes with any Bayesian model that gives priors over functions such as deep Gaussian processes \citep{damianou2013deep} or Bayesian neural networks \citep{goan2020bayesian}.

With the rise in lab automation, this workflow can be integrated into a design build test pipeline similar to \cite{carbonell_automated_2018} and \cite{hamedirad_towards_2019} which can greatly reduce the time required to optimize new biomolecular components, speeding up the creation of new devices. This method could also be incorporated into hybrid models in bio-processing and chemical engineering, for decision making for systems with many similar components  \citep{narayanan_hybrid_2023, mowbray_machine_2021, schweidtmann_machine_2021}.

\section{Conclusion}

We have shown how a transfer learning design of experiments workflow can be used to optimize many competitor DNA molecules for an amplification-based diagnostics device. We used cross-validation to demonstrate that the latent variable multi-output Gaussian process has the best predictive accuracy and have shown it has the least regret when Bayesian optimization is performed on the DNA amplification data. Future improvements to the optimization of the model hyperparameters would lead to faster and more consistent performance of the algorithm. Despite this, we believe this workflow is applicable to many other biotechnology applications and should be used to reduce the experimental load when there are many similar tasks to be optimized but their similarity is a priori unknown.

\newpage

\bibliography{references.bib}
\clearpage
\newpage

\appendix

\section{Nomenclature} \label{supp:nom}

\renewcommand{\nomname}{}
\printnomenclature
\label{supp:nom}

\nomenclature[A]{PCR}{Polymerase Chain Reaction}    
\nomenclature[A]{BP}{Number of Base Pairs}    
\nomenclature[A]{GC}{Percentage Guanine-Cytosine Content}    
\nomenclature[A]{DNA}{Deoxyribonucleic Acid}    
\nomenclature[A]{MOGP}{Multi-output Gaussian Process}    
\nomenclature[A]{LMC}{Linear Model of Coregionalization}    
\nomenclature[A]{LVMOGP}{Latent Variable Multi-output Gaussian Process} 

\nomenclature[A]{RMSE}{Root Mean Squared Error}    
\nomenclature[A]{NLPD}{Negative Log Predictive Density}    
\nomenclature[A]{AvgGP}{Average Gaussian Process}    

\onecolumn

\section{Latent Variable Multi-output Gaussian Process Implementation} \label{supp:LVMOGP_VI}

Gaussian processes are normally trained by maximizing the log marginal likelihood. However, the presence of the latent variable distributions in the LVMOGP means the log marginal likelihood is no longer tractable. Instead, \cite{dai_efficient_2017} used variational inference to approximate a lower bound to this log marginal likelihood, following the method proposed by \cite{titsias_variational_2009} and \cite{titsias_bayesian_2010}. In variational inference, the aim is to minimize the Kullback-Leibler divergence between an approximate posterior and a true posterior. 

Our implementation of the LVMOGP takes a concatenation of the input data and their corresponding latent variables \(\Tilde{X} = [X, H_:] \in \mathbb{R}^{N \times (D + Q_H)} \) where \(H_:\) to denotes the vector of latent inputs for each observed data point. All inputs \(X_p\) for the same output dimension will have the same latent variable, \(\boldsymbol{h}_p\). 

For the LVMOGP this variational lower bound is given as:

\begin{multline}
ELBO =  \sum_{i=1}^N 
\left[ -\frac{1}{2\sigma_n^2}y_{:i}^Ty_{:i}
+\frac{1}{\sigma_n^2}y_{:i}\langle K_{iu}\rangle_{q(H_{:i})}K_{uu}^{-1}M
 \right. \\ \left. - \frac{1}{2\sigma_n^2} Tr(K_{uu}^{-1} \langle K_{iu}^TK_{iu} \rangle_{q(H_{:i})} K_{uu}^{-1}(MM^T+S)) \right. \\ \left.
 - \frac{1}{2\sigma_n^2} ( Tr(\langle K_{ii}\rangle_{q(H_{:i})}) - Tr(K_{uu}^{-1} \langle K_{iu}^TK_{iu}\rangle_{q(H_{:i})}) \right] \\  -\frac{NP}{2} log(2\pi\sigma_n^2) - KL[q(\mathbf{u}) || p(\mathbf{u})] - \sum^{N}_{i=1} KL[q(H_{:i})||p(H_{:i})],
\end{multline}
\noindent
where \(\langle K\rangle_{q(h_i)}\) denotes a kernel expectation over the variational distribution of the latent variable of data point \(i\). \(K_{ii}\) and \(K_{uu}\) are the covariance functions of the data and the inducing points \(Z\) respectively, while \(K_{iu}\) is the cross covariance function between the two. \(\mathit{Tr}\) is the trace of a matrix. \(M\) and \(S\) are the mean and covariance of the variational distribution over inducing points \(q(Z) \sim \mathcal{N}(M, S)\). The second term in this expression can be viewed as a data fit term, while the last term can be seen as a complexity penalty. 

\nomenclature[A]{ELBO}{Evidence lower bound to marginal likelihood for LVMOGP} 
\nomenclature[F]{\(K_{ii}(\cdot, \cdot)\)}{Covariance function of the data} 
\nomenclature[F]{\(K_{uu}(\cdot, \cdot)\)}{Covariance function of the inducing points} 
\nomenclature[F]{\(K_{iu}(\cdot, \cdot)\)}{Cross covariance function between the data and inducing points} 
\nomenclature[P]{\(Z\)}{Inducing points} 
\nomenclature[P]{\(M\)}{Mean of the variational distribution on Z} 
\nomenclature[P]{\(S\)}{Variance of the variational distribution on Z} 
\nomenclature[P]{\(q(\cdot)\)}{Variational distribution} 
\nomenclature[P]{\(\mathbf{u}\)}{Inducing variables} 

Two types of prediction are relevant using the LVMOGP. The first is when we have new input points \(X_*\) and new position on the latent space \(h_*\). In this case, the posterior prediction can be calculated in closed form. The second, and more likely, prediction case is when we want to predict a new point \(X_*\) at a point on the latent space where we already have data with latent variable \(h_{p}\). This integration is intractable, but following \cite{titsias_bayesian_2010}, the first and second moments can be computed in closed form if using a squared exponential kernel. 

\section{Toy Dataset Creation} \label{app:toy_dataset}

The dataset used in Figure~\ref{fig:model_comparison} was generated by creating two latent functions from samples of a Gaussian process prior with the squared exponential kernel, in Equation~\ref{squared_exp}, and \(\ell = 0.3\) and \(\sigma_k^2 = 2\) and multiplying them by random weights to create the output functions. To ensure the output functions could generate data anywhere, a Gaussian process was fitted to the densely sampled points for each output function. Data was then generated by evaluating the mean of the Gaussian processes at varying input locations adding noise \(\epsilon = \mathcal{N}(0, 0.1)\). Similarly to the competitor dataset, the amount and location of data observed on each output function varies.

\section{Gaussian Process Implementation} \label{app:GPR}

\subsection{Data Standardization} \label{app:datastand}
For both the synthetic and competitor datasets we standardize the input and output data by subtracting the mean and dividing by the standard deviation, such that:
\begin{equation}
   \bar{x} = \frac{x-\mu_x}{\sigma_x},
\end{equation}
and similar for the output data. This is common practice for Gaussian process regression as it reduces numerical instability and allows for better interpretability of hyperparameter values, which is useful for initialization.

\subsection{Choice of Gaussian Process Prior} \label{app:GPprior}
When using Gaussian process models for real-world optimization tasks the Gaussian process prior should be informed by existing knowledge of the system. For example, the choice of kernel function can express belief of the smoothness or periodicity of the function and a mean function may be selected if there is a known trend in the data \citep[Chapters 2 \& 4]{rasmussen_gaussian_2006}. For the competitor design task, we believed the function to be smooth, so opted for the squared exponential kernel in Equation~\ref{squared_exp}. We did not have any prior information about a trend in the data so used a zero mean function.

\subsection{Gaussian Process Hyperparameter Training} \label{supp:initalizations}

Gaussian processes are generally trained using the marginal likelihood, which automatically trades off data fit and model complexity, guarding against overfitting \citep[Chapter 5]{rasmussen_gaussian_2006}. Ideally, we would perform full Bayesian inference over the Gaussian process hyperparameters, however, this is often difficult and expensive due to the need to use approximation techniques to evaluate intractable integrals \citep{lalchand2020approximate}.

Instead, we use type II maximum likelihood, a common approach of maximizing the marginal likelihood with respect to the hyperparameters. With sufficient data, this approach is justified based on the Laplace approximation and because in practice the posterior for the hyperparameters tends to be highly peaked \citep{mackay1999comparison}.
However, at lower data regimes, the non-convexity of the marginal likelihood surface can cause overfitting due to multiple modes \citep{lalchand2020approximate}. Low data regimes can also lead to hyperparameters being weakly identified, leading to flat ridges in the marginal likelihood surface, making the optimization sensitive to starting values \citep{warnes1987problems}. So, at low data regimes, Gaussian processes trained with type II maximum likelihood can over fit. As more data is collected, the type II maximum likelihood approach is a reasonable approximation and the marginal likelihood will automatically trade off model complexity and data fitting, preventing overfitting \citep[Chapter 5]{rasmussen_gaussian_2006}.

The marginal likelihood optimization surface is non-convex and therefore gradient based optimizers will only find local optima meaning the result dependent on initialization \citep{mackay_introduction_1998}.

To overcome this, and reduce overfitting, we use random restarts, along side principled methods of initialization, to fit the same Gaussian process model multiple times, and then select the hyperparameter configuration with the best log marginal likelihood. These regimes, introduced in Appendix~\ref{app:hyp_init}, differ slightly for the different models. We use gradient descent to optimize the marginal likelihood each time.

\subsection{Hyperparameter Initialization} \label{app:hyp_init}

For all model, unless otherwise states, we initialize the lengthscale randomly as \(\ell \sim \text{Uniform}(0,1)\), noise variance randomly as \(\sigma_n \sim \text{Uniform}(0,0.1)\) which is equivalent to the noise being between \(0\%\) and \(10\%\) of the data variance and kernel variance \(\sigma_k = 1\). These settings are standard proactive for Gaussian process regression \citep{matthews_gpflow_2017}. For the MOGP and AvgGP we did nine random restarts with these settings. 

For the LMC we used three different methods for initializing \(\mathbf{W}\) and \(\mathbf{\kappa}\), with three random restarts for each:
\begin{itemize}
    \item \emph{Both \(\mathbf{W}\) and \(\mathbf{\kappa}\) random}. In this initialization, we initialize
    \(\mathbf{W}\sim\text{Uniform}(0.1, 1)\) and  \(\mathbf{\kappa}\sim\text{Uniform}(0.1, 1)\).
    \item \emph{\(\mathbf{W}\) random and \(\mathbf{\kappa}=0\)}. In this initialization 
    \(\mathbf{W}\sim\text{Uniform}(0.1, 1)\) and \(\mathbf{\kappa}=10^{-6}\). This initialization was chosen as we thought it would favor solutions with small \(\mathbf{\kappa}\) so it would better fit the linear correlation case, where the test functions are generated as linear combinations of some linear functions. 
    \item \emph{\(\mathbf{W}\) random and \(\mathbf{\kappa}=1\)}. In this initialization 
    \(\mathbf{W}\sim\text{Uniform}(0.1, 1)\) and \(\mathbf{\kappa}=1\). We chose this initialization to favor large \(\mathbf{\kappa}\), which is useful for the uncorrelated test case, as it would encourage the output functions to behave independently of each other. 
\end{itemize}

The random initialisations for \(\mathbf{W}\) helped the initialisations for two reasons: firstly, in the GPflow implementation if  \(\mathbf{W}\) is not initialized it defaults to a rank of 1, and secondly by initializing to random values rather than all one value we avoid saddle points on the optimization surface. 

For the LVMOGP we used three different initialization procedures, again with three random restarts for each:
\begin{itemize}
    \item \emph{Random}. In this initialization all hyperparameters and variational parameters were initialized randomly. the means of the latent variables were initialized as \(\mu_H \sim \text{Uniform}(-1, 1)\).
     \item \emph{GPy}. This is the method used in the GPy implementation of the LVMOGP \citep{dai_efficient_2017}, that has following three steps:
     \begin{enumerate}
        \item A sparse MOGP is fitted to the data using a set of inducing points \(Z\) which are common to all outputs. The mean predictions \(\mu(Z) \in \mathbb{R}^{N_U \times P}\) of the output function values at these inducing inputs is then calculated:
        \begin{equation}
            \mu(Z) = K(Z, Z)[K(Z, Z) + \sigma_n^2\mathbf{I}]^{-1}Y.
        \end{equation}
        
        The sparse MOGP is used is ensure all output functions are observed at the same input locations for the functional PCA, which is necessary when data is observed at different locations on different surfaces. It also serves the purpose of smoothing the data plus the trained lengthscales are used to initialise the lengthscales of the observed dimensions of the LVMOGP. 
    
        \item The mean predictions \(\mu(Z) \in \mathbb{R}^{N_U \times P}\) are then used as inputs to functional PCA. 
        The first \(Q_H\) eigenvectors \(V \in \mathbb{R}^{N_U \times Q_H}\) and eigenvalues \(\{\lambda_q\}_{q=1}^{Q_H}\) of \(\mu(Z)^T \mu(Z)\) are calculated and used to project \(\mu(Z) \) into latent space
        \begin{equation}
            H = \mu(Z)^T V,
        \end{equation}
    
        where \(H \in \mathbb{R}^{P \times Q_H}\). The relative contributions of each of the eigenvalues is also calculated as: 
        \begin{equation}
            \varsigma_q = \frac{ \tilde{\lambda}_q}{\max\{\tilde{\lambda}_i\}_{i=1}^{Q_H}} ~~~~ \tilde{\lambda}_q = \frac{\lambda_q}{\sum_{i=1}^{Q_H} \lambda_i}
        \end{equation}
    
        \item The latent variables \(H\) from the functional PCA are used to initialize the latent variables of a Bayesian Gaussian process latent variable model. The lengthscales of the Bayesian Gaussian process latent variable model are initialized to \(\{\frac{1}{\varsigma_q}\}_{q=1}^{Q_H}\). Once the Bayesian Gaussian process latent variable model is trained, the latent variables and hyperparameters of the Bayesian Gaussian process latent variable model are used to initialize those of the LVMOGP. 
    \end{enumerate}
    
    \item \emph{PCA}. In this initialization, the first two steps of the \emph{GPy} initialization are followed. This means fitting a sparse MOGP to the data and performing principle component analysis (PCA) on the posterior predictions at inducing point locations. The MOGP hyperparameters were then used to initialize the LVMOGP observed lengthscale, kernel variance and noise variance. The output of the PCA was used to initialize the latent variable means and the lengthscale of the latent dimensions. This initialization was chosen as a simplified version of the \emph{GPy} initialization.
   
\end{itemize}

See the github repositories in Appendix~\ref{supp:github} for more details.

In the synthetic experiments, we found the method of initializing the hyperparameters affected the end log marginal likelihood, with no initialization outperforming all others for each model. Therefore, we decided to continue with all initializations for the PCR data experiments. For the PCR data experiments we did 10 random restarts for each initialization, due to the randomness of some of the initializations. 

\subsection{Latent Dimensions} \label{app:latent_dims}
The LMC and LVMOGP have hyperparameters that need to be set for the number of latent functions and dimensions respectively.

For the LMC, the rank of the similarity matrix \(\mathbf{B}\) needs to be selected. This is equivalent to the number of latent functions \citep{alvarez_kernels_2012}. If the rank is too low, the LMC will fail to explain the data well. However if it is too high, the LMC can suffer from overfitting when data is limited. Unlike other hyperparameters, such as the lengthscale or noise variance, there is no continuous way to select this hyperparameter. Therefore, to select this parameter for a given problem, it is necessary to fit multiple LMC models with different ranks and select the best one by comparing the marginal likelihoods or cross validation. In reality, this isn't always feasible as it is computationally expensive, and data may be limited.

\begin{figure}
    \centering
    \includegraphics[width=12cm]{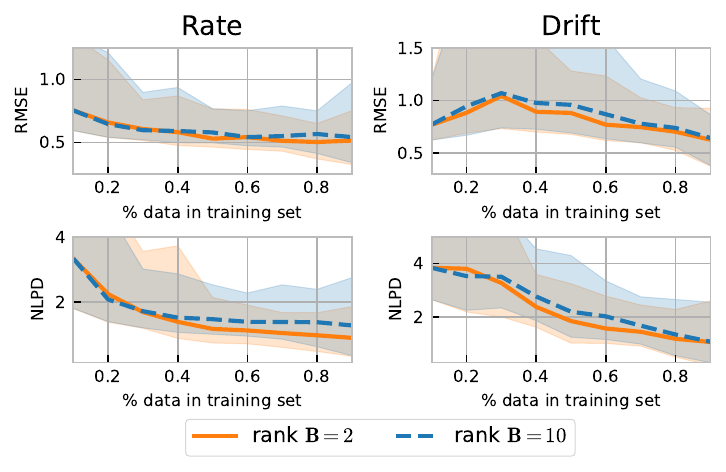}
    \caption{Cross validation results for the linear model of coregionalization with similarity matrices \(B\) with rank \(2\) and \(10\). For each percentage train, \(70\) different train test splits were used. The LMC with rank \(10\) matrix has worse NLPD than that with rank \(2\), suggesting overfitting is occurring.}
    \label{fig:xvalid_diff_latents_lmc}
\end{figure}

Figure~\ref{fig:xvalid_diff_latents_lmc} shows results of cross validation experiments introduced in Section~\ref{sec:cross_val} for the LMC  with this latent dimension hyperparameters set to \(2\) and \(10\). The test setting was the same as in Section~\ref{sec:cross_val} and the cross validation was repeated \(70\) times. The LMC has worse NLPD  with more latent functions, most likely because more hyperparameters have been introduced, increasing the chances of overfitting when the dataset isn't large.

Similarly, for the LVMOGP, the dimensionality of the latent space needs to be selected. However, if a kernel that treats each dimension as independent is used, the LVMOGP can "switch off" unessential dimensions \citep{titsias_bayesian_2010}. This is done by making the lengthscales of the unessential dimensions really large, so there is no variation across those dimensions. This means the LVMOGP can automatically reduce the number of dimensions to those that give a good trade off between data fit and model complexity. This effect occurs in the latent dimensions of the drift parameter in Figure~\ref{fig:LVMOGP_latent_space}, where all the points are lined up on a single dimension.

\begin{figure}
    \centering
    \includegraphics[width=12cm]{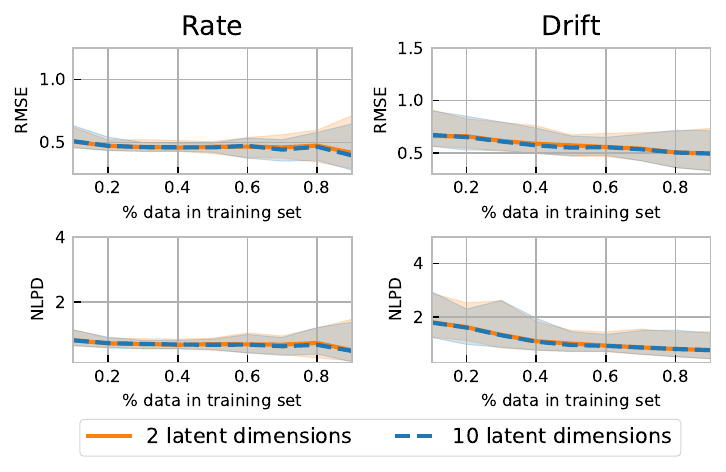}
    \caption{Cross validation results for the latent variable multioutput Gaussian process with \(2\) and \(10\) dimensional latent spaces. For each percentage train, \(70\) different train test splits were used. Due to the automatic relevance determining properties of the latent space kernel, there is very little difference in performance.}
    \label{fig:xvalid_diff_latents_lvm}
\end{figure}

Figure~\ref{fig:xvalid_diff_latents_lvm} shows results of cross validation experiments introduced in Section \ref{sec:cross_val} for the LVMOGP with \(2\) and \(10\) latent dimensions. Unlike the LMC, the LVMOGP however performs the same as it can switch off unnecessary dimensions for the \(10\) latent case.

Generally when conducting Bayesian optimization, data is very scarce to begin with, so it is not possible to perform model selection for the rank of the LMC \(\mathbf{B}\) matrix. One option would be to fit many models with different ranks and select the one with the best marginal likelihood, but this is generally too computationally expensive. This is one of the limitations of the LMC. Therefore, for the Bayesian optimization experiments we used a rank of \(2\) for the LMC and \(2\) dimensions for the LVMOGP as we expected these settings to give enough flexibility to transfer information while limiting chances of overfitting.

\section{Data and Code Availability} \label{supp:github}

Raw data is available on request from rdm-enquiries@imperial.ac.uk.

Implementation of the methods outlined in this paper requires a basic knowledge of python. We provide two github repositories, which include the methods used and jupyter notebooks demonstrations.
The first repository, \url{https://github.com/RSedgwick/TLGPs}, contains code for each of the Gaussian process models, the synthetic experiments and some jupyter notebooks demonstrating the use of the models. It also contains instructions for running the code.
This repository is application agnostic and could easily be transferred to other use cases.

The second repository, \url{https://github.com/RSedgwick/TL_DOE_4_DNA}, is specifically tailored to the competitor use case. This repository contains the code for the cross validation and Bayesian optimization experiments, as well as code for data processing and notebooks for results analysis. This repository is more targeted to our specific use case. 

\section{Data Summary} \label{data_summary}

Each competitor is defined by its primer-reporter combination. For each of these primer-pair combinations we then have data at different guanine-cytosine  content and no. of base pairs combinations. Table~\ref{tab:data_summary} gives a summary of the number of unique locations on each of the competitors. 

\begin{table}[t!]
\centering
\begin{tabular}{|c|>{\centering\arraybackslash}p{2.1cm}|c|>{\centering\arraybackslash}p{2.1cm}|}
\hline
\multicolumn{2}{|c|}{Not To Be Optimized} & \multicolumn{2}{c|}{To Be Optimized} \\ \hline
Primer Reporter Combination & No. Unique Locations & Primer Reporter Combination & No. Unique Locations \\ \hline
FP004-RP004-EvaGreen      &   28 & FP004-RP004-Probe       & 53 \\
FP002-RP002x-Probe        &   12 & FP001-RP001x-EvaGreen   & 24 \\
FP004-RP004x-Probe        &   12 & FP001-RP001x-Probe      & 20 \\
FP001-RP001-Probe         &    9 & RP001x-FP002-Probe      & 19 \\
FP001-RP005-Probe         &    8 & FP002-RP002x-EvaGreen   & 15 \\
FP004-RP004x-EvaGreen     &    8 & FP005-FP001-EvaGreen    & 14 \\
FP003-RP008-Probe         &    5 & FP004-FP005-Probe       &  8 \\
FP006-RP006-Probe         &    5 & FP005-FP001-Probe       &  8 \\
FP005-RP005-Probe         &    5 & FP005-FP004-EvaGreen    &  8 \\
FP002-RP002-EvaGreen      &    4 & RP002x-FP005-Probe      &  8 \\
FP002-RP006-Probe         &    4 & RP008x-FP001-EvaGreen   &  8 \\
FP057.1.0-RP003x-Probe    &    3 & RP008x-FP005-Probe      &  8 \\
FP003-RP008x-EvaGreen     &    3 & FP001-RP004-EvaGreen    &  7 \\
FP003-RP008-EvaGreen      &    3 & RP002x-FP004-EvaGreen   &  6 \\
FP002-RP002-Probe         &    3 & FP002-RP004-EvaGreen    &  3 \\
FP001-RP001-EvaGreen      &    2 & RP002x-FP002-EvaGreen   &  2 \\
FP003-RP003-Probe         &    1 & &  \\
FP057.1.0-RP003x-EvaGreen &    1 & &  \\ \hline

\end{tabular}
\caption{Summary of the amount of data we have for each competitor design surface. Each unique location refers to a unique GC-BP combination.}
\label{tab:data_summary}
\end{table}
\vspace{2mm}

\section{Extra Bayesian Optimization Results} \label{supp:bayes_opt}

The following tables contain extra results for the Bayesian optimization experiments. The first table in each section, Tables \ref{tab:best_point_unconstrained} and \ref{tab:best_point_constrained}, shows counts of the first model to get to the best point on a surface for all competitors and seeds. If two models get to the best point on the same iteration, they are both counted as "winners". The second table, Tables  \ref{tab:low_cum_unconstrained} and \ref{tab:low_cum_constrained} shows counts of the models with the lowest cumulative regret for each competitor and seed. The same thing applies if two models have the same cumulative regret. 
For the single objective optimization, Table~\ref{tab:iter_best_unconstrained} shows the average number of iterations for each model to get within tolerance of the target rate (+/- 0.05). For the penalized optimization Table~\ref{tab:iter_best_constrained} shows the average number of iterations for each model to get either within tolerance of the rate target with no drift penalty, or to the best point (which may have a drift penalty).  For some of the runs with the drift penalty, some of the models failed to get to the best point for some surfaces within the experimental budget. In these cases, those surfaces were discarded and the average was taken for the surfaces where all the models had managed to get to the best point within the experimental budget.

\vspace{2mm}

\subsection{Single Objective Optimization} \label{app:single_obj}

Extra results for the single objective Bayesian optimization. These results demonstrate that the LVMOGP gets to the best point more often (Table~\ref{tab:best_point_unconstrained}) and has has the lowest cumulative regret (Table~\ref{tab:low_cum_unconstrained}) more often than the other models. The LVMOGP also reaches the best point in the lowest number of iterations for all the learning scenarios (Table~\ref{tab:iter_best_unconstrained}). 

\begin{table*}[h!]
    \centering    
    \begin{tabularx}{\textwidth}{|>{\centering\arraybackslash}m{3cm}|>{\centering\arraybackslash}p{3cm}|>{\centering\arraybackslash}p{2cm}| >{\centering\arraybackslash}p{2cm}| >{\centering\arraybackslash}p{2cm} | >{\centering\arraybackslash}p{2cm}|}
    \cline{1-6}
     learning scenario & starting point &  MOGP & Avg GP  & LMC & LVMOGP \\ \cline{1-6}
    \multirow{2}{7em}{learning many} & center & 124  & 121  & 144 & \textbf{255} \\ \cline{2-6}
       & model's choice & 107 & 119  & 97 & \textbf{147}  \\ \cline{1-6}
       \multirow{2}{6em}{one at a time} & center & 140 & 140 & 156 & \textbf{215}   \\ \cline{2-6}
       & model's choice & 86 & 118 & 87 & \textbf{191} \\ \cline{1-6}
    \end{tabularx}
    \caption{Table showing counts of the first Gaussian process model to reach the best point on a surface for the single objective Bayesian optimization experiments. The counts are the number of times a Gaussian process model did the best on a competitor for each seed. If two Gaussian process models performed the same for a given instance, they are both counted. This is for 16 competitors and 25 random seeds.}
    \label{tab:best_point_unconstrained}
\end{table*}
\begin{table*}[h!]
    \centering
       \begin{tabularx}{\textwidth}{|>{\centering\arraybackslash}m{3cm}|>{\centering\arraybackslash}p{3cm}|>{\centering\arraybackslash}p{2cm}| >{\centering\arraybackslash}p{2cm}| >{\centering\arraybackslash}p{2cm} | >{\centering\arraybackslash}p{2cm}|}
    \cline{1-6}
     learning scenario & starting point &  MOGP & Avg GP  & LMC & LVMOGP \\ \cline{1-6}
    \multirow{2}{7em}{learning many} & center & 182  & 80  & 140 & \textbf{197}\\ \cline{2-6}
       & model's choice & 85 & 94 & 83 & \textbf{117}  \\ \cline{1-6}
       \multirow{2}{6em}{one at a time} & center & 129 & 140 & 131 & \textbf{206}  \\ \cline{2-6}
       & model's choice & 99 & 106 & 87& \textbf{159} \\ \cline{1-6}
    \end{tabularx}
    \caption{Table showing counts of the Gaussian process model with the lowest cumulative regret on a surface for the single objective Bayesian optimization experiments. The counts are the number of times a Gaussian process model did the best on a competitor for each seed. If two Gaussian process models performed the same for a given instance, they are both counted.  This is for 16 competitors and 25 random seeds.}
    \label{tab:low_cum_unconstrained}
\end{table*}

\begin{table*}[h!]
    \centering

      \begin{tabularx}{\textwidth}{|>{\centering\arraybackslash}m{3cm}|>{\centering\arraybackslash}p{3cm}|>{\centering\arraybackslash}p{2cm}| >{\centering\arraybackslash}p{2cm}| >{\centering\arraybackslash}p{2cm} | >{\centering\arraybackslash}p{2cm}|}
    \cline{1-6}
     learning scenario & starting point &  MOGP & Avg GP  & LMC & LVMOGP \\ \cline{1-6}
    \multirow{2}{7em}{learning many} & center &  3.13 & 3.25 & 3.11 &  \textbf{2.58} \\ \cline{2-6}
       & model's choice & 3.08 & 2.63 & 3.09 & \textbf{2.44} \\ \cline{1-6}
       \multirow{2}{6em}{one at a time} & center & 2.94 &  3.06 & 2.85 & \textbf{2.15}  \\ \cline{2-6}
       & model's choice & 2.94 & 2.63 &  2.63 &  \textbf{1.81}  \\ \cline{1-6}
    \end{tabularx}
    \caption{Table showing the mean number of iterations need for the models to get within tolerance of the target rate (+/- 0.05) for the single objective optimization. This is for 16 competitors and 25 random seeds.}
    \label{tab:iter_best_unconstrained}
\end{table*}

\vspace{4mm}
\subsection{Bayesian Optimization with Drift Penalty} \label{app:bayes_opt_with_drift}
\vspace{4mm}
Extra results for the Bayesian optimization with a penalty on drift. These results demonstrate that the LVMOGP gets to the best point more often (Table~\ref{tab:best_point_constrained}) and has has the lowest cumulative regret (Table~\ref{tab:low_cum_constrained}) more often than the other models for most of the learning scenarios. The LVMOGP also reaches the best point in the lowest number of iterations for all the learning scenarios (Table~\ref{tab:iter_best_constrained}). 

\begin{table*}[h!]
    \centering
    
      \begin{tabularx}{\textwidth}{|>{\centering\arraybackslash}m{3cm}|>{\centering\arraybackslash}p{3cm}|>{\centering\arraybackslash}p{2cm}| >{\centering\arraybackslash}p{2cm}| >{\centering\arraybackslash}p{2cm} | >{\centering\arraybackslash}p{2cm}|}
    \cline{1-6}
    learning scenario & starting point &  MOGP & Avg GP  & LMC & LVMOGP \\ \cline{1-6}
      learning scenario & starting point &  MOGP & Avg GP  & LMC & LVMOGP \\ \cline{1-6}
    \multirow{2}{7em}{learning many} & center &  142 & 157 & 123 & \textbf{165} \\ \cline{2-6}
       & model's choice &  89 & \textbf{122} & 101 & 111  \\ \cline{1-6}
       \multirow{2}{6em}{one at a time} & center & 141 &  137 & 153 & \textbf{217}   \\ \cline{2-6}
       & model's choice & 75 & 102  & 79 & \textbf{164}\\ \cline{1-6}
    \end{tabularx}
    \caption{Table showing counts of the first Gaussian process model to reach the best point on a surface for the penalized Bayesian optimization experiments. The counts are the number of times a Gaussian process model did the best on a competitor for each seed. If two Gaussian process models performed the same for a given instance, they are both counted. This is for 16 competitors and 24 random seeds.}
    \label{tab:best_point_constrained}
\end{table*}

\begin{table*}[h!]
    \centering
    
      \begin{tabularx}{\textwidth}{|>{\centering\arraybackslash}m{3cm}|>{\centering\arraybackslash}p{3cm}|>{\centering\arraybackslash}p{2cm}| >{\centering\arraybackslash}p{2cm}| >{\centering\arraybackslash}p{2cm} | >{\centering\arraybackslash}p{2cm}|}
    \cline{1-6}
     learning scenario & starting point &  MOGP & Avg GP  & LMC & LVMOGP \\ \cline{1-6}
    \multirow{2}{7em}{learning many} & center & \textbf{180} & 118 & 100 & 163 \\ \cline{2-6}
       & model's choice & 85 & 103 & 84 &  \textbf{111} \\ \cline{1-6}
       \multirow{2}{6em}{one at a time} & center & 173 &  118 & 139  & \textbf{204}  \\ \cline{2-6}
       & model's choice & 83 & 70 & 65 &  \textbf{156}  \\ \cline{1-6}
    \end{tabularx}
    \caption{Table showing counts of the Gaussian process model that had the lowest cumulative regret on a surface for the penalized Bayesian optimization experiments. The counts are the number of times a Gaussian process model did the best on a competitor for each seed. If two Gaussian process models performed the same for a given instance, they are both counted.  This is for 16 competitors and 24 random seeds.}
    \label{tab:low_cum_constrained}
\end{table*}

\begin{table*}[h!]
    \centering
    
      \begin{tabularx}{\textwidth}{|>{\centering\arraybackslash}m{3cm}|>{\centering\arraybackslash}p{3cm}|>{\centering\arraybackslash}p{2cm}| >{\centering\arraybackslash}p{2cm}| >{\centering\arraybackslash}p{2cm} | >{\centering\arraybackslash}p{2cm}|}
    \cline{1-6}
     learning scenario & starting point &  MOGP & Avg GP  & LMC & LVMOGP \\ \cline{1-6}
    \multirow{2}{7em}{learning many} & center &  \RS{3.56} & \RS{4.03}  & \RS{3.96} & \RS{\textbf{3.47}} \\ \cline{2-6}
       & model's choice & \RS{3.78} & \RS{3.00} & \RS{3.66} & \RS{ \textbf{2.63}}  \\ \cline{1-6}
       \multirow{2}{6em}{one at a time} & center & \RS{3.70}  & \RS{\textbf{3.44}}  & \RS{3.51}  &  \RS{3.56} \\ \cline{2-6}
       & model's choice & \RS{3.63}  & \RS{3.06}  & \RS{3.46} & \RS{\textbf{2.52}} \\ \cline{1-6}
    \end{tabularx}
    \caption{Table showing the mean number of iterations need for the models to either get within tolerance of the target rate (+/- 0.05) without drift penalty or reach the best point (which may have a penalty) for the penalized optimization. For some runs, one or more of the models would not achieve this within the experimental budget. In these cases, the number of iterations to best point was set to the experimental budget. This is for 16 competitors and 24 random seeds.}
    \label{tab:iter_best_constrained}
\end{table*}

\subsection{Comparison of Choice of First Point} \label{supp:first_point}

Table~\ref{tab:first_points} shows the average regret of the first data point chosen by each of the models for each of the learning scenarios, for the single objective case. From this table, it is clear to see the AvgGP and the LVMOGP improve on the regret of the central point, and outperform the random selection of the MOGP and LMC. This demonstrates that having a principled method of selecting the first point is useful for reducing regret. 

\begin{table*}[h!]
    \centering
    
    \begin{tabularx}{\textwidth}{|>{\centering\arraybackslash}m{3cm}|>{\centering\arraybackslash}p{3cm}|>{\centering\arraybackslash}p{2cm}| >{\centering\arraybackslash}p{2cm}| >{\centering\arraybackslash}p{2cm} | >{\centering\arraybackslash}p{2cm}|}
    \cline{1-6}
     learning scenario & starting point &  MOGP & Avg GP  & LMC & LVMOGP \\ \cline{1-6}
    \multirow{2}{7em}{learning many} & center & 0.588 & 0.588   & 0.588 & 0.588  \\ \cline{2-6}
       & model's choice & 0.651 & 0.499  & 0.703 & 0.464  \\ \cline{1-6}
       \multirow{2}{6em}{one at a time} & center & 0.588 & 0.588   & 0.588 & 0.588 \\ \cline{2-6}
       & model's choice & 0.675 & 0.308 & 0.623 & 0.309 \\ \cline{1-6}
    \end{tabularx}
    \caption{Table of the mean regret of the first data point for each of the learning scenarios for each of the models.}
    \label{tab:first_points}
\end{table*}

\subsection{Comparison of Number of Experiments} \label{app:no_experiments}

\begin{figure}
    \centering
    \includegraphics[width=0.9\textwidth]{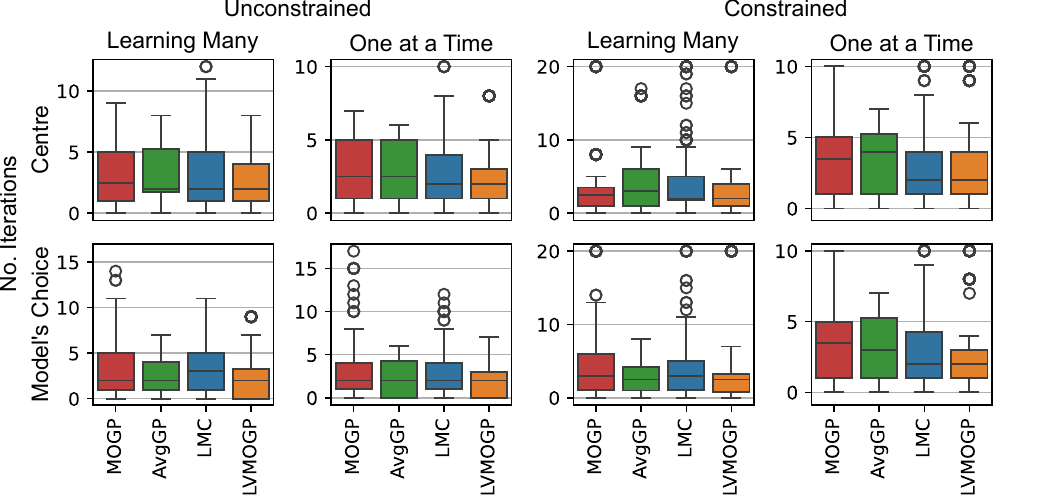}
    \caption{Box plots of the number of iterations needed to reach within \(0.05\) tolerance of the best point for all learning scenarios for the Bayesian optimization experiments outlined in Section~\ref{sec:bayes_opt_res}.}
    \label{fig:iterations_boxplot}
\end{figure}

Figure~\ref{fig:iterations_boxplot} shows box plots of the number of iterations taken by each model to reach the best point for each learning scenario for the experiments in Section~\ref{sec:bayes_opt_res}. The distributions here are across the \(16\) different competitors and \(24\) random restarts. For the constrained optimization case, for some seeds some of the models didn't reach the best point for competitor FP004-RP004-Probe within the experimental budget. In these cases we set the number of experiments to the total experimental budget (\(10\) or \(20\) depending on the scenario).

These results show the LVMOGP on average requires less experiments than the other models to select the best point, demonstrating how this approach can reduce the number of experiments needed to optimize the competitors. All the models have some outliers, this could be due to being unlucky with the initial data it receives or sub optimal hyperparameter optimization.

\subsection{Initial Surfaces for Bayesian Optimization} \label{app:init_surfaces}

In the Bayesian optimization experiments outlined in section ~\ref{sec:bayes_opt_res}, we start with the two competitor surfaces with the most data fully observed, FP004-RP004-EvaGreen and FP002-RP002x-Probe. This is so the Gaussian process models have the chance to learn a reasonable prediction before the iterative Bayesian optimization process. It also replicates the case where we already have limited data on a couple of competitors and want to optimize more competitors. 

\begin{figure}
    \centering
    \includegraphics[scale=0.9]{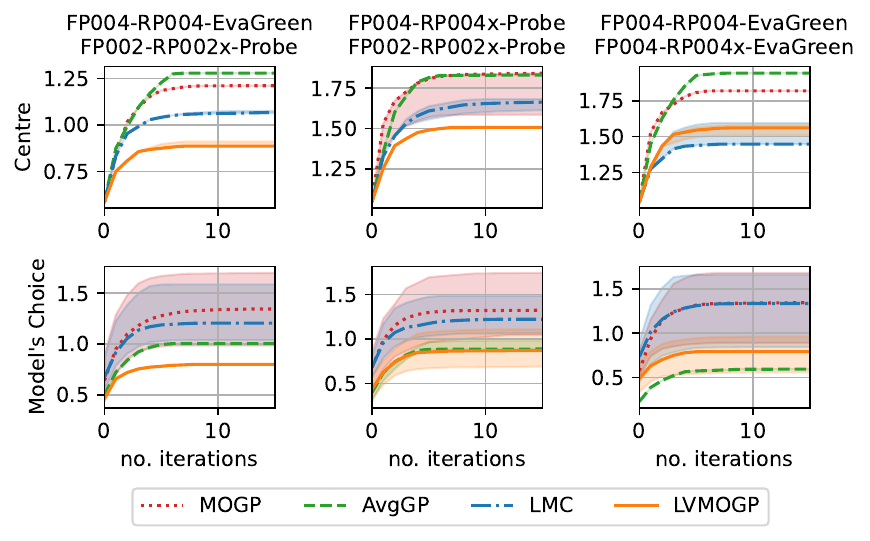}
    \caption{Results of single objective Bayesian optimization experiments for the \textit{learning many} scenario with different initial two competitors. This experiment was run 20 times with different seeds.}
    \label{fig:diff_inits_single}
\end{figure}

\begin{figure}
    \centering
    \includegraphics[scale=0.9]{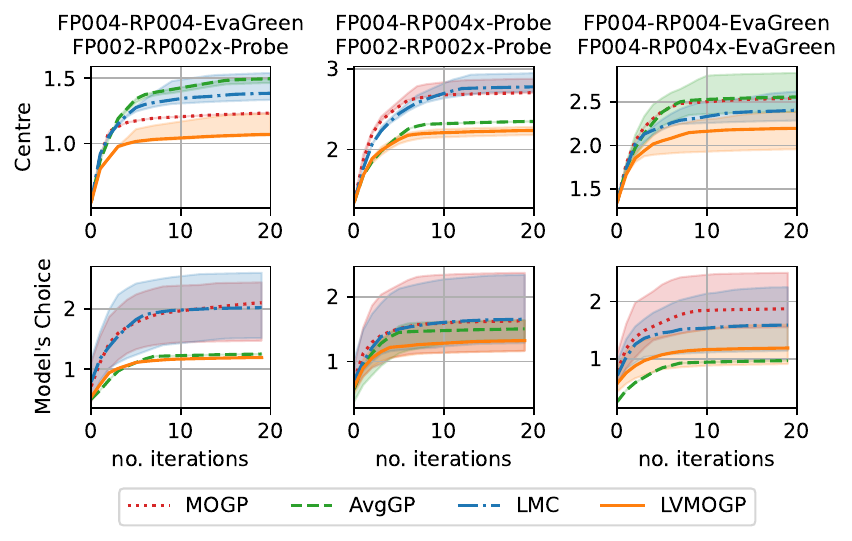}
    \caption{Results of penalized objective Bayesian optimization experiments for the \textit{learning many} scenario with different initial two competitors. This experiment was run 20 times with different seeds.}
    \label{fig:diff_inits_penalized}
\end{figure}

To assess whether the choice of initial surfaces affects the results in our experiments, we investigate two other combinations of two initial surfaces. We ran the \textit{learning many} scenario of Bayesian optimization with \(20\) seeds for both the single objective and penalized optimization. We chose the  \textit{learning many} as we expect the initial surfaces to have more impact in this scenario than in the \textit{one at a time} scenario. The results of these experiments are plotted in Figures~\ref{fig:diff_inits_single} and \ref{fig:diff_inits_penalized}. In these results, it is clear the initial surfaces make some difference to the models' performances, although the LVMOGP still performs the best most of the time. The differences in performance are most likely due to different amounts of initial data (depending on how much data we have for the initial surfaces) and differences in the similarity of the initial surfaces to the surfaces to be learned.

\section{Funding Information}

This work was supported by the UKRI CDT in AI for Healthcare Grant No. EP/S023283/1, UK Research and Innovation Grant No. EP/P016871/1, the BASF / RAEng Research Chair in Data-Driven Optimization, the US NIH Grant No. 5F32GM131594, the EPSRC IRC Next Steps Plus grant No. EP/R018707/1 and the RAEng Chair in Emerging Technologies award No. CiET202194. For the purpose of open access, the authors have applied a Creative Commons Attribution (CC BY) licence to any Author Accepted Manuscript version arising.

\section{Author Contributions}

\noindent
J.G. conducted lab experiments. R.S. developed code, conducted code experiments and wrote the manuscript. R.M. and M.v.W. supervised the project, specifically giving guidance on the machine learning aspects. J.G. and M.S. also supervised the project, specifically giving guidance on the bioengineering aspects.

\end{document}